\documentstyle[12pt]{article}
\def\ben{\begin{enumerate}}  \def\een{\end{enumerate}}
\def\beq{\begin{equation}}   \def\eeq{\end{equation}}

\def\MSbar{\relax\ifmmode\overline{\rm MS}\else{$\overline{\rm MS}${ }}\fi}
\def\gbar{\relax\ifmmode\overline{g}\else{$\overline{g}${ }}\fi}
\def\albar{\overline \alpha}  
\def\define{\buildrel \rm def \over =}
\def\gbar{\overline g}     \def\etc{\hbox{\it etc.}{ }}
\begin{document}
\begin{titlepage}
\thispagestyle{empty}
\begin{flushright}
{\small \today \hfill Dedicated to the memory of Nicolaj N. Bogoliubov  \\
on the occassion of his 90th birthday.}
\end{flushright}
\vspace{18mm}

\begin{center}
{\large\bf  Evolution of the Bogoluibov Renormalization Group}
\vspace{3mm}

\author{D.V.~Shirkov} 
\vspace{2mm}

  {\it N.N.Bogoliubov Laboratory, JINR, Dubna, Russia; \\
   shirkovd@thsun1.jinr.ru}
\end{center}
\vspace{12mm}

\abstract{ We start with a simple introduction into the renormalization
group (RG) in quantum field theory and give an overview of the
renormalization group method. The third section is devoted to essential
topics of the renorm-group use in the QFT.  Here, some fresh results are
included.

  Then we turn to the remarkable proliferation of the RG ideas into various
fields of physics. The last section  summarizes an impressive recent
progress of the ``QFT renormalization group" application in mathematical
physics.}

\vspace{15mm}
\end{titlepage}

\pagenumbering{roman}
\tableofcontents
\newpage
\pagenumbering{arabic}

\section{Renormalization group primer \label{s1}}
\subsection{Mathematical preliminaries \label{ss1.1}}
\subsubsection{\small\it Renorm-group folklore}

Let us start with some simple statements which can be supposed to be widely
known by particle theorists. In the quantum field theory (QFT) the
renormalization group (RG) is usually associated with a possibility of
presenting any physical quantity, $F(Q^2, g),$~ calculated under a definite
renormalization prescription in the form $F(Q^2/\mu^2, g_\mu)~$ (for
simplicity -- in a massless case) with the renormalized coupling constant
$g_\mu$ definition attached to some renormalization point (or reference
momentum scale) $Q=\mu$.  Differential RG equation is usually said to be
driven from the condition that $F$ does not depend on the choice of $\mu$~,
\beq \label{1-1} {dF\over d\mu} = 0 ~~. \eeq

  The coupling constant $g_{\mu}$ dependence on $\mu$ is described by a
 specific function $\gbar(Q^2)$ known as an {\it effective coupling}
(sometimes -- effective coupling constant) $g_{\mu}=\gbar (\mu ^2)~$.

Eq.(\ref{1-1}) can be written down in the form of a partial linear
differential equation (DE)
\beq\label{1-2}
 \left[x{\partial\over{\partial x}} - \beta(g){\partial\over
{\partial g}}\right] ~F(x, g) = 0 ~ \eeq
where $x=Q^2/\mu^2,~g$ stands for $g_{\mu}$ and $\beta (g)$, the group
generator, usually referred to as {\it beta function}, is defined by
$$
\beta(g_{\mu}) = z{\partial \gbar(z) \over \partial z}
~~~~~\hbox{ at }~~~z = \mu^2~. $$

 The effective coupling  $\,\gbar\,$ should be considered as a function
of two arguments: $x= Q^2/\mu^2\,$ and $\,g_\mu\,$~ with the boundary
condition $\,\gbar(1,g)=g$. Besides
\beq\label{1-3}
 \left[x{\partial\over{\partial x}} - \beta(g){\partial\over
{\partial g}}\right] \gbar(x, g) = 0 ~ \eeq
it satisfies the nonlinear DE
\beq\label{1-4}
x{{\partial\gbar(x, g)}\over {\partial x}} = \beta(\gbar(x, g)) ~ \eeq
which is nothing else but a characteristic equation for (\ref{1-2}). To
employ this formalism, one has to give $~\beta(g)$. Usually, for
this one uses renormalized perturbation theory. \par
\smallskip

The foregoing can be considered as a ``RG folklore". For brevity, we
gave it in the simplest {\it massless} version, which corresponds to the UV
case, for the QFT model with one coupling constant.

\subsubsection{\small\it Group Functional Equation}

 Less popular are the RG Functional Equations (FEs). The FE for the
$\gbar$  in the UV case has the form
\beq\label{1-5}
\gbar(x, g) = \gbar\left( ~\frac{x}{t}~, ~~\gbar(t, g)\right)~. \eeq
 This equation, which follows (see, e.g., the Chapter {\it Renormalization
Group} in Ref.\cite{Book}) from finite Dyson renormalization transformations,
represents a basement of the differential RG formulation. Popular DE
(\ref{1-4}) can be directly obtained from it by differentiating over $x$ and
then putting $t=x$. On the other hand, by differentiating (\ref{1-5}) with
respect to $t$ at $t=1$ we get partial DE (\ref{1-3}).

  The FE (\ref{1-5}) as well as similar FEs for propagators and vertex
 functions (see, below, eq.(\ref{feqs})) must be considered as the most
 compact and general formulation of the RG symmetry in QFT. \par

  However, in reality, group FEs, like (\ref{1-5}) (and DEs (\ref{1-4}) and
(\ref{1-3}) as well) do not contain any physics at all being just the
reflection of the group composition law! Here, we mean the continuous group
(that is, the Lie group of transformations) of operations changing the
reference point $\mu$ involved into  the coupling constant $g_{\mu}$
definition. Namely, we can regard the change of a reference coupling
$g_{\mu} \to g_{\bar{\mu}} $ as an operation of the group element $T_t$
$$
T_t g_{\mu }=g_{\mu \sqrt{t}}= \gbar (t,g_{\mu })~$$
with a real
continuous positive numerical parameter $t~\/(=\mu^2/\bar{\mu}^2)~$.

If we set $x=\tau t$, then the l.h.s. of (\ref{1-5}) can be achieved from $g$
by operation $T_{\tau t}$, while the r.h.s. may be identified as $T_{\tau}
T_t\/g$. The content of eq.(\ref{1-5}) is just the group composition law,
$$
T_{\tau t} = T_\tau T_t ~.$$

Thus, the essence of the basic RG functional equation~(\ref{1-5}) is
the necessary condition for transformations $T_t$ to form a group. \par

At the same time, it demonstrates that function $\bar g$ is {\it invariant}
with respect to simultaneos transformation
\beq\label{1-10}
R_t~:~~\{~ x'=x/t~,~~g'=\bar{g}(t,g)\}~. \eeq
Invariance condition for an observable now can be written down as
$$
F(x, g) = F\left( ~\frac{x}{t}~, ~~\gbar(t, g)\right)~.$$

 Usually, of interest are also functions $\phi(x,g)$ (like, e.g., propagator
amplitudes in QFT) transforming as a linear representation of RG
\beq\label{1-11}
\phi (x,g) ~\to~R_t \phi = \phi (x',g') = z(t, g)\phi(x,g) ~.\eeq

 Note also that the group FE for an observable, like matrix element, is
of the form
\beq\label{1-18}
M(\{x\},y;g) =M\left(\left\{\frac{x}{t}\right\},\frac{y}{t},
g(t,y;g)\right) ~;~~ \{x\}= x_1, x_2, \dots, x_k ~ \eeq
which reflects an existence of several $Q^2$-type arguments and implements
its independence of renormalization details corresponding to Eq.(\ref{1-1}).

\subsubsection{\small\it Abstract Formulation}

 To make this point clearer, let us show that FE (\ref{1-5}) can formally be
obtained directly from the group composition law. Generally, the
mathematical formulation of the RG transformation can be presented as a
functional realization of the mentioned Lie group. \par

  Consider the transformation $T(l)$ of a certain abstract set ${\cal M}$
of elements $M_{i}$ into itself depending on a continuous real parameter
$l~~(-\infty < l < \infty)$ such that for each element $M$ we have
$$
T(l) M = M^\prime ~~(M, M^\prime \subset {\cal M})~.$$
Suppose that ${\cal M}$ can be projected onto the real axis, i.e., for every
$M_i$ there corresponds a real number $g_i$\footnote{This condition is not
essential and can be modified --- see, below, eqs. (\ref{1-16}) and
(\ref{1-17}).}. Then, this transformation can be written in the analytic form
$$
T(l)g = g'= G(l,g)~,$$
$G$ being a continuous function of two arguments satisfying the
normalization condition $G(0,g) = g$ which corresponds to the identity
transformation $T(0) = {\bf E}$. \par

   Transformations $T(l)$ form a group if they satisfy the composition law
\beq\label{composit}                                              
T(l)\times T(\lambda ) = T(l+\lambda ) \eeq                       
to which there corresponds the functional equation for $G$ :
\beq\label{1-7}
G\{ l, G(\lambda, g)\} = G ( l +\lambda, g)~.\eeq                              
  As it follows from the bases of the Lie group theory, it is sufficient to
deal with the infinitesimal transformation at $\lambda\ll 1$ , i.e., with
the DE
\beq\label{1-8}
{\partial G(l,g)\over \partial l} = \beta \{ G (l,g) \}~.\eeq     
 Here the group generator is defined as
$$
\beta (g) = {\partial G (\epsilon ,g)\over \partial \epsilon }
~~~\mbox{at}~~~\epsilon = 0 .$$

\par Performing a logarithmic change of variables
\begin{equation}\label{log}
l = \ln x,~~~\lambda = \ln t~,~~~~ G(l,g) = \bar{g}(x,g)~, ~~T(\ln t)=T_t
\end{equation}
we obtain (\ref{1-5}) and (\ref{1-4}) instead of (\ref{1-7}) and (\ref{1-8}).

\subsection{Definition of the Renorm-Group \label{ss1.2}}
\subsubsection{\small\it The RG transformation}

 Generally, the RG can be defined as a continuous one-parameter
group of specific transformations of a partial solution (or the solution
characteristic) of a problem, a solution that is fixed by a boundary
condition. The RG transformation involves boundary condition parameters and
corresponds to some change in the way of imposing this condition. \par

 For illustration, imagine an one-argument solution characteristic $f(x)$
that has to be specified by the boundary condition $f(x_{0}) = f_{0}$.
Formally, represent the given characteristic of a partial solution as a
function of boundary parameters as well:  $f(x) = f(x,x_{0},f_{0})$. (This
step can be considered as an {\it embedding} operation). The RG
transformation then corresponds to a changeover of the way of
parameterization, say from $\{x_{0},f_{0}\}$ to $\{x_1,f_1\}$ for the {\it
same} solution.  In other words, the $x$ argument value, at which the
 boundary condition is given, does not need to be $x_0$, but we may choose
another point $x_i$. Our solution $f$ can be written in a form of a
two-argument function $F(x/x_0,f_0)$ with the property $~F(1,\gamma)=\gamma$.
The equality $F({x/x_0}, f_0) = F({x/x_1}, f_1)$ reflects the fact that under
such a change of a boundary condition the form of  function $F$ itself is
not modified (as, e.g., in the case of $~F(x,\gamma)=\Phi(\ln x +\gamma)$).
Noting that $~f_1 = F(x_1/x_0, f_0)~$, we obtain
$$
F(\xi, f_0) = F(\xi/t, F(t,f_0)) ~~;~~~\xi = x/x_0~, ~t = x_1/x_0~. $$
The group transformation here is $\{~\xi\to\xi/t,~~f_0 \to f_1=F(t,f_0)~\}~.$
\medskip

{\sf The renorm-group transformation} for a given solution of some physical
problem in the simplest case can be defined as \par
\smallskip

{\it a simultaneous one-parameter transformation of two variables,} say $x$
and $g$, by
$$
R_t~~:~~ \{~x\to x' = x/t~,~~g\to g'=\gbar(x,g)~ \}~,\eqno{(6)} $$
\smallskip

\noindent the first being a scaling of a coordinate $x$ and the second --- a
more complicated functional transformation of the solution characteristics.
Eq.(\ref{1-5}) for the transformation function $\gbar$ provides the group
property $T_{\tau t} = T_\tau T_t ~$  of the transformation (6).  Performing
the logarithmic change of variables and an appropriate redefinition of a
transformation function (\ref{log}), we obtain eqs.(\ref{1-7}), (\ref{1-8})
and
\beq\label{1-9}
R(l)~~:~~ \{~ q\to q' = q - l~,~~ g\to g' = G(l,g)~ \}~,\eeq
\smallskip
instead of (\ref{1-5}), (\ref{1-4}) and (6). One can refer to these equations
as the {\it multiplicative} version (and previous equations in abstract
formulation as to the {\it additive} one). They are just the RG equations and
transformation for a massless QFT model with one coupling constant. In that
case $x =Q^{2}/\mu^{2}$ is the ratio of a 4-momentum $Q$ squared to a
``normalization" momentum $\mu $ squared and $g$, the coupling constant.
\par

Several generalizations of (\ref{1-10}) and (\ref{1-9}) will be considered
below.

\subsubsection{\small\it Simple generalizations \label{sss1.2.2.}}

 {\sf ``Massive" Case}

  For example in QFT, if we do not neglect the mass $m$ of a particle, we
have to insert an additional dimensionless argument into the invariant
coupling $\bar{g}$ which now has to be considered as a function  of three
variables:  $x=Q^{2}/\mu ^{2},~y=m^{2}/\mu ^{2}$, and $g$. The presence of a
new ``mass" argument $y$  modifies the group transformation
\beq\label{1-12}
R_t~:~~\left\{x'=\frac{x}{t}~,~~y'=\frac{y}{t}~,~~g'=\bar{g}(t,y;g)~\right\}
\eeq 
 and the functional equation
\beq\label{1-13}
\bar{g}(x,y;g) =\bar{g}\left(~\frac{x}{t}~,~\frac {y}{t}~;~
\bar{g}(t,y;g)\right)~. \eeq                                   
 Here, it is important that the new parameter $y$ (which in physical nature
must be close to the $x$ variable as it scales similarly) enters also
into the transformation law of $g$ . \par
  If the considered QFT model, like QCD, contains several masses there will
be several mass arguments
$$
y\to\{y\}=y_{1}, y_{2}, \ldots y_{n}~. $$
\smallskip

{\sf Multi-coupling case}

  A more complicated generalization corresponds to transition to the
case with several coupling constants: $g\to \{g\} = g_{1},\ldots g_{k}~.$
Here, one has to introduce the ``family" of effective couplings
$$ 
\bar{g}\to\{\bar{g}\}~,~~\bar{g}_i=\bar{g}_i(x,y;\{g\})~;~~~~~i=1,2,\ldots k~,
$$  
\noindent satisfying the system of coupled functional equations
\beq\label{1-16}                                                  
\bar{g}_{i}(x,y;\{g\})=\bar{g}_{i}\left(\frac{x}{t}~,~\frac{y}{t}~;~
\{ ~\bar{g}(t,y; \{g\} )~\}~\right)~. \eeq        
In the abstract formulation this system is a generalization of (\ref{1-5})
and (\ref{1-13}) for the case when every element $M_{i}$ of ${\cal M}$ can be
described by $k$ numerical parameters, i.e., by a point $\{ g\}$ in the
$k$-dimensional real parameter space. The RG transformation now is
\beq\label{1-17}
R_t~:~\left\{~x \to \frac{x}{t}~,~~y \to \frac{y}{t}~,~~\{g\} \to
\{g(t)\}~\right\}~~~,~~ g_i(t)=\gbar_i(t,y;\{g\})~.  \eeq         

\subsection{Early history and the RG method \label{ss1.3} }       

\subsubsection{\small\it Renormalization and renormalization invariance}
  As it is known, the regular formalism for eliminating the UV divergences in
 QFT was developed on the basis of covariant perturbation theory for the
scattering $S$--matrix in the late 40s. This breakthrough is connected with
the names of Tomonaga, Feynman, Schwinger and some others. In particular,
Dyson and Abdus Salam carried out the general analysis of the structure of
divergences in arbitrarily high orders of perturbation theory. Nevertheless,
a number of subtle questions concerning overlapping divergences remained
unclear.

  An important contribution in this direction based on a thorough analysis of
the mathematical nature of UV divergences was made by Bogoliubov. This was
achieved on the basis of a branch of mathematics which was new at that time,
namely, the Sobolev--Schwartz~ {\it theory of distributions}. The point is,
that propagators in local QFT are distributions (similar to the Dirac
delta--function) and propagator products appearing in the coefficients of the
$S$--matrix expansion require a supplementary definition in the
case when their arguments coincide and lie on the light cone.

 In the mid 50s on the basis of this approach Bogoliubov and his disciples
developed a technique of supplementing the definition of products of
singular St\"uckelberg--Feynman propagators \cite{ufn55} and proved a
theorem~\cite{paras} on the finiteness and uniqueness (for renormalizable
theories) of the $S$--matrix elements in any order of perturbation theory.
The prescriptive part of this theorem, the {\it Bogoliubov R-operation} (see,
e.g., chapter ``Removal of divergencies from the $S$--matrix" in the
 monograph \cite{Book}), still remains a practical means of obtaining finite
and unique results in the higher order perturbation calculation.
 \smallskip

  The Bogoliubov algorithm works, essentially, as follows: \par
-- To remove the UV divergences of a one-loop diagram, instead of introducing
some regularization, e.g., the momentum cutoff, and handling (quasi)
infinite counterterms, it suffices to complete the definition of a divergent
Feynman integral by subtracting from it a certain polynomial in the external
momenta which in the simplest case is reduced to the first few terms of the
Taylor series.

-- For multi-loop diagram (including one with overlapping divergences) one
 should first subtract all divergent subdiagrams and finish with subtracting
the diagram as a whole in a hierarchical order regulated by the
 $\,R$--operator. \par
  \smallskip

 An attractive feature of this approach is that it is free from any auxiliary
nonphysical attributes such as bare masses, bare coupling constants, and
regularization parameters which are not involved in the computation within
the Bogoliubov's algorithm. The latter can be regarded as {\it
renormalization without regularization and counterterms}.

 The uniqueness of computational results for the observable $S$-matrix
elements is ensured by special conditions imposed on them. These conditions
contain some degree of freedom (related to different renormalization schemes
and momentum scales) that can be used to establish finite relations
 between the Lagrangian parameters (masses, coupling constants) and
corresponding renormalized quantities. The fact that physical predictions are
independent of arbitrariness in the renormalization conditions, that is, they
are {\it renorm--invariant}, constitutes the conceptual foundation of the
renormalization group.

\subsubsection{\small\it The discovery of the renormalization group
\label{sss1.3.2}}                                                  
 In the 1952-1953 St\"uckelberg and Peterman \cite{stp} discovered\footnote{
For a more detailed exposition of the RG early history see our review
\cite{brown93}.} a group of infinitesimal transformations related to finite
arbitrariness arising in the $S$-matrix elements upon elimination of the UV
divergences. These authors introduced {\it normalization group} generated by
Lie operators connected with renormalization of the coupling constant $e$.

 In the following year, on the basis of (infinite) Dyson's renormalization
transformations formulated in the regularized form, Gell-Mann and
Low~\cite{gml} derived functional equations for the QED propagators in the UV
limit.  The appendix to this article contains the general solution (obtained
by T.D. Lee) of this functional equation for the renormalized transverse
photon propagator amplitude $d(x,e^2)$, written in two equivalent forms:
\begin{equation}\label{1-19}
e^2d\left(x,e^2\right)=F\left(xF^{-1}\left(e^2\right)\right)~~~\mbox{and}
 \;~~~\ln x=\int\limits_{e^2}^{e^2d}\frac{{\rm d}y}{\psi(y)}~,
~~\psi(e^2)=\left.\frac{\partial(e^2d)}{\partial\ln x}\right|_{x=1}~.
\end{equation}
A qualitative analysis of the behaviour of the quantum electromagnetic
interaction at small distances was carried out with the aid of (\ref{1-19}).
Two possibilities, namely, infinite and finite charge renormalizations were
pointed out.

However, paper~\cite{gml} paid no attention to the group character of the
analysis and the results obtained there. The authors missed a chance to
establish a connection between their results and the standard perturbation
theory and did not discuss the possibility that a ghost pole solution might
exist.

 The final step was taken by Bogoliubov and the present author
{bs-55a,bs-55b,sh-55} --- see also the survey~\cite{nc-56} published in
English in 1956. Using the group properties of finite Dyson transformations
for the coupling constant, fields and Green functions, these authors derived
functional group equations for the propagators and vertices in QED in the
general case (that is, with the electron mass taken into account). For
example, the equation for the transverse amplitude of the photon propagator
and electron propagator amplitude were obtained in the form
$$
d(x,y;e^2)=
d(t,y;e^2)~d\left(\frac{x}{t},\frac{y}{t};~e^2d(t,y;e^2)\right),\;$$
\begin{equation} \label{feqs}
s(x,y;e^2)=s(t,y;e^2)~ s\left(\frac{x}{t},
\frac{y}{t};~ e^2d(t,y;e^2)\right)  \end{equation}
in which the dependence on the mass variable $y=m^2/\mu^2$ was present.

 As can be seen, the product  $e^2d\,$ of electron charge squared and photon
propagator amplitude enters in both the FEs. This product is invariant with
respect to Dyson's transformation. We called this function -- {\it invariant
charge} and introduced the term {\it renormalization group}. \par

 In the modern notation, the first equation is that for the invariant
charge (now widely known as an effective or running coupling)
$\bar\alpha =\alpha d(x,y;\alpha=e^2)$:
\begin{equation}    \label{1-21}
 \bar\alpha(x,y;\alpha)=
\bar\alpha\left(\frac{x}{t}, \frac{y}{t}; ~\bar\alpha(t,y;\alpha)\right)~.
\end{equation}

 Let us emphasize that, unlike the approach Ref.\cite{gml}, in the latter
case there is no relation with UV divergences and simplification due to
the massless nature of the UV asymptotics.  Here, the homogeneity of the
transfer momentum scale is violated explicitly by  mass $m$. Nevertheless,
the symmetry (even though a bit more complex one) underlying the RG, as
before, can be stated as an {\it exact symmetry} of the solutions of the QFT
problem.  This is what we mean when using the term {\sf Bogoliubov's
renormalization group} or {\it renorm-group} for short.  \smallskip

The differential group equations for $\bar\alpha$ and for the electron
propagator:
\begin{equation}\label{1-22}
\frac{\partial\bar\alpha(x,y;\alpha)}{\partial\ln x}=
\beta\left(\frac{y}{x},\bar\alpha(x,y;\alpha)\right)\,\,;\,\,
\frac{\partial s(x,y;\alpha)}{\partial\ln x}=
\gamma\left(\frac{y}{x},\bar\alpha(x,y;\alpha)\right)s(x,y;\alpha)~, \eeq
with
\begin{equation} \label{1-23}
\beta(y,\alpha)=\frac{\partial\bar\alpha(\xi,y;\alpha)}{\partial\xi}~,~
~~~\gamma(y,\alpha)=\frac{\partial s(\xi,y;\alpha)}{\partial\xi}~~~~
\mbox{at}~~\xi=1~ \end{equation}
were first derived in \cite{bs-55a} by differentiating the FEs. In this way,
explicit realization of the group DEs mentioned in the paper \cite{stp} was
obtained. These results established a conceptual link with the
St\"uckelberg---Peterman and Gell-Mann---Low results.

\subsubsection{\small\it Creation of the RG method\label{sss1.3.3}} 

Another important achievement of paper \cite{bs-55a} consisted in formulating
a simple algorithm for improving an approximate perturbative solution by
combining it with the Lie equations (for detail, see below, Section 2).

 In our adjacent publication~\cite{bs-55b} this algorithm was effectively
used to analyse the UV and infrared (IR) behaviour in QED. The
one-loop and two-loop UV asymptotics
\begin{equation} \label{a1rg}
\bar\alpha^{(1)}_{RG}(x;\alpha)\equiv \bar\alpha^{(1)}_{RG}(x,0,\alpha)=
\frac{\alpha}{1-\frac{\alpha}{3\pi}\ln x}\,\,,  \end{equation}
\begin{equation}   \label{a2rg}
\bar\alpha^{(2)}_{RG}(x;\alpha)= \frac{\alpha}{1-\frac{\alpha}{3\pi}\ln x
+\frac{3\alpha}{4\pi}\ln(1-\frac{\alpha}{3\pi}\ln x)}
\end{equation}
of the photon propagator as well as the IR behavior
\beq\label{ir}
s(x,y;\alpha)\approx (x/y-1)^{-3\alpha/2\pi}=(p^2/m^2-1)^{-3\alpha/2\pi}~\eeq
of the electron propagator in the transverse gauge were obtained. At that
time, these expressions were already known only at the one--loop level. It
should be noted that in the mid 50s the problem of the UV behaviour in local
QFT was quite urgent. At that time, substantial progress in the analysis of
QED at small distances was made by Landau and his collaborators~\cite{lakh}.
However, Landau's approach did not provide a prescription for constructing
subsequent approximations.

  The simple technique for obtaining higher approximations was found only
within the new renorm--group method. The one-loop UV asymptotics of QED
propagators obtained in our paper~\cite{bs-55b}, eqs. (\ref{a1rg}) and
(\ref{ir}), agreed precisely with the results of Landau's group.

  Within the RG approach these results can be obtained in just a few lines
of argumentation. To this end, e.g., the massless one-loop perturbation
approximation should be substituted into the r.h.s. of the
first equation in (\ref{1-23}) to compute the generator
$\beta(0,\alpha)=\psi(\alpha)= \alpha^2/3\pi$ followed by elementary
integration of the first of eqs.(\ref{1-22}).

  Moreover, starting from the next order perturbation expression
 $\bar\alpha^{(2)}_{PTh}(x,;\alpha)$ containing the $\alpha^3\ln x$ term, we
arrived at the second renorm-group approximation (\ref{a2rg}) performing
summation of the $\alpha^2(\alpha\ln)^n$ terms. This two-loop solution for
the invariant coupling first obtained in~\cite{bs-55b} contains the
nontrivial log--of--log dependence which is now widely known of the
``next-to-leadind logs" approximation for the running coupling in quantum
 chromodynamics (QCD) --- see, below, eq.(\ref{2-23}).

 Comparing solution (\ref{a2rg}) with (\ref{a1rg}), one can conclude that
the two-loop correction is extremely essential just in the vicinity of the
ghost pole singularity at $\,x_1=\exp{(3\pi/\alpha)}$. This demonstrates that
the RG method is a regular procedure, within which it is quite easy to
estimate the range of applicability of the results.

Quite soon, this approach was formulated~\cite{sh-55} for the case of QFT
with two coupling constants, say, $g$ and $h$, namely, for a model of
pion--nucleon interactions with the pions self-interaction. To the system of
functional equations for two invariant couplings
\begin{eqnarray} \label{}
&&\bar g^2\left(x,y;~ g^2,h\right)=
\bar g^2\left(\frac x t,~ \frac y t,~ \bar g^2(t,y; g^2, h),
~\bar h\left(t,y;~g^2, h\right)\right)~,\nonumber\\
&&\bar h\left(x,y;~g^2,h\right)=
\bar h\left(\frac x t,~ \frac y t, ~\bar g^2\left(t,y;~g^2,h\right),
~\bar h\left(t,y;~g^2, h\right)\right)~\nonumber
\end{eqnarray}
there corresponds a coupled system of nonlinear DEs -- see, below,
eqs.(\ref{3-5}). It was analysed \cite{ilya} in the one-loop appriximation
to carry out the UV analysis of the renormalizable model of the pion-nucleon
interaction.

 In a more general case of arbitrary covariant gauge the RG analysis in QED
was carried out in~\cite{tolia}.  Here, the point was that the charge
renormalization is connected only with the transverse part of the photon
propagator. Therefore, under nontransverse covariant (e.g., Feynman) gauge
the Dyson transformation has a more complex form.  This issue has been
resolved by considering the gauge parameter as another coupling constant.

 In Refs.\cite{bs-55a,bs-55b,sh-55,tolia} and \cite{ilya} the RG was thus
directly connected with practical computations of the UV and IR asymptotics.
Since then this technique, known as the {\sl renormalization group method}
(RGM) and being summarized in the first edition of monograph \cite{Book},
has become the sole means of asymptotic analysis in local QFT.

\subsection{RG in QED \label{ss1.4}}                              
\subsubsection{\small\it  Effective Electron Charge}

 An essential feature of quantum theory is the presence of virtual states and
transitions. In QED, e.g., the process of virtual dissociation of a
photon into an electron-positron pair and vice versa $\gamma \leftrightarrow
e^{+}+e^{-}$ can take place. The sequence of two such virtuale transitions
represents the simplest contribution to the effect of vacuum polarization.

  The vacuum polarization processes lead to several specific phenomena and
particularly to the notion of {\it effective electron charge}. To explain
this, let us start with a classical analogy.  \par

  Take a polarizable medium consisting of molecules that can be imagined as
electric dipoles. Insert into it an external electric charge $Q$.  Due to the
attraction of opposite charges, the dipoles change their position so that the
charge $Q$ turns out to be partially screened. As a result, at a distance $r$
from $Q$ the electric potential will be smaller than the vacuum Coulomb law
$Q/r$ and can be presented in the form $Q(r)/r$ where, generally, $Q(r)\le
Q$. The introduced quantity $Q(r)$ is known as an effective charge.  As $r$
decreases, $Q(r)$ increases and as $r\to 0,~Q(r)$ tends to {\it Q}.

  In QFT the vacuum, i.e., the interparticle space itself stands for the
``polarizable  medium". Quantum--field vacuum is not physically empty. It is
filled with vacuum fluctuations, i.e., with virtual particles. These ``zero
fluctuations" are a well-known effect of a ground state in quantum world. In
QED, zero oscillations consist mainly of short-lived virtual $(e^{+},e^{-})$
pairs which play the role of tiny electric dipoles.

 Consider the process of measuring the electron charge with the help of some
external electromagnetic field. In the quantum case the probing photon can
virtually dissociate into the $(e^+,e^-)$ pair. This pair can be treated as a
virtual dipole that produces partial screening of the measured charge. The
simplest process involves two elementary electromagnetic interactions, its
contribution to an effective charge being proportional to the small number
$e^{2} \equiv \alpha ~\simeq ~1/137$; and this contribution depends on the
distance $r$ !  In the region of $r$ values much smaller than the Compton
length of the electron $r_e=h/mc~\simeq ~3,9.10^{-11} cm$ it depends on
 $r$ logatithmically
\begin{equation}\label{dir}
e~\to ~e(r)=e \left\{1-{\alpha\over 3\pi}~\ln {r\over r_{e}}+\dots\right\}
\end{equation}
 as was first discussed by Dirac \cite{dirac} in the middle of the 30s. The
$e(r)$ value decreases as $r$ grows. So, qualitatively, the QED effective
charge behavior corresponds to a classical picture of screening.

 This dependence can be presented by a set of curves $e(r)$.  Each curve
represents a possible behavior of the effective charge $e(r)$ as obtained
from the theory and considered without any reference to experiment
($\alpha=e^{2}$ being unspecified numerically).

 The point is that in the classical analogue the value of an external charge
$Q$ inserted into the polarizable medium is known from the very beginning. In
quantum physics it is not the case, and a charge value can be measured at
{\it not very small} distances. The result of measurement generally has to be
specified by two quantities: the ``distance of measurement" $r_{i}$ and the
measured charge value $e_{i}$. Hence, to make the choice from the mentioned
set of curves, one has to fix the point on the plane with the coordinates
$\,r=r_i,~ e(r)=e_i$. Thus, for the chosen ``physical" curve $\,e(r_i)=e_i$.
Note, that the usual definition of the electron charge by a classical
macroscopic (like Millikan) experiment corresponds here to very large
distances $r~\geq~r_{e}$, i.e., $ e=e(r= r_e )= 1/\sqrt{137}$.  \par

As it is well known, in relativistic microphysics one usually uses the
momentum rather than coordinate representation.  Correspondingly, instead of
$e(r)$ one deals with the quantity $\bar{\alpha}(Q^2)$, the Fourier transform
of $e(r)$ squared. It is a monotonically increasing function of its argument
$Q^{2}$, the 4-momentum transfer squared. Here and below, the bar denotes a
function (distinct from $\alpha,~\alpha _{\mu },~\alpha _{i}$ -- its
numerical values at some given value of the $Q^{2}$ argument). The
correspondence condition with the classical electrodynamics now takes the
form $\bar{\alpha}(0)=1/137\/$, as in our scale the external long-range field
corresponds to a photon with vanishing 4-momentum. However, as before, to fix
one of possible curves on the plane~$(Q,~\bar{\alpha})$ one has to give a
point $Q\equiv \sqrt{Q^2}=\mu,~\bar{\alpha}=\alpha_{\mu}\/$ and hence, for
the selected curve $~\bar{\alpha}(\mu^2)= \alpha_{\mu}\/$.

  The parameter~$\mu$ sometimes is referred to as a {\it scale parameter}.
As is clear, it is just the momentum magnitude for a photon used for the
charge measurement. The effective coupling function $\bar{\alpha}(Q^2)$
describes the dependence of the electron charge value on the measurement
conditions. In our days the logarithmic corrections to the Millikan value
become essential and are measured at big accelerators.  \par

  The parameter $\mu$ has no analogue in the QED Lagrangian that reproduces
the classical electrodynamical one. The phenomenon of its arising in QFT was
exaggerated by the term ``dimensional transmutation". As it was shown, its
appearance is very natural and is connected with the measurement procedure.

 This is a good place to recall the ideas by Niels~Bohr formulated in the
middle of 30s \cite{bohr} and related to the complementarity principle.
The point is that to specify a quantum system, it is necessary to fix its
``macroscopic surrounding", i.e., to give the properties of macroscopic
devices used in the measurement process.  Just these devices are described by
additional parameters, like $\mu\/$.  However, this is not the end of the
Bohr (i.e. -- scale) parameter story. As can be shown, in the QFT this
parameter existence leads to a new symmetry lying in the foundation of the
renormalization group.

\subsubsection{\small\it RG transformations}                            

 To do this, consider again the function $\bar{\alpha}(Q^2)$ having in mind
that the physical solution has been chosen by the condition
$\bar{\alpha}(Q^2=\mu^2)=\alpha_{\mu}$.  Assume also for simplicity that we
deal with a massless QED,~more precisely, with the approximation $\mid Q\mid
\gg $~{\it m}. This corresponds to the GeV-energy region or to distances
$r\ll r_{e}$.  Here the effective charge function ~$\bar{ \alpha}$ can be
represented as a function of two dimensionless arguments $Q^2/\mu^2$ and
$\alpha_{\mu}$, i.e., $\bar{\alpha}(Q^2)~=~\bar{\alpha
}(Q^2/\mu^2,\alpha_{\mu})~.$

 Now, take into account that the couple of parameters $\mu, \alpha_{\mu}$
used to identify the physical solution may, generally, correspond to any
scale $\mu$.  Take two scales ``1" and ``2"  with  coordinates
$\/\mu_1, \alpha_1$ and $\mu_2, \alpha_2\/$, respectively. It is evident that
$\bar{\alpha}$ can be parameterized by any pair $\/\mu_i, \alpha_i\/;
~i=1,2,\ldots~$ so that for arbitrary $Q^2$ values the identity
$$
\bar{\alpha}(Q^2/\mu^2_1, \alpha_1)=\bar{\alpha}(Q^2/\mu^2_2, \alpha_2)$$
should hold. \par

  At the same time the second argument in the r.h.s., $\alpha_2$, which by
definition is equal to $\bar{\alpha}$ at $Q^2=\mu_2^2$, can be expressed in
terms of $\bar{\alpha}$ parameterized with the help of point "1" coordinates,
i.e., $\alpha_2=\bar{\alpha}(Q^2=\mu_2^2)=\bar{\alpha}(\mu_2^2/\mu_1^2,
\alpha_1)~.$ Combining the last two relations and introducing a notation
$Q^2/\mu^2=x,~~\alpha_1=\alpha,~~\mu_2^2/\mu_1^2=t~,$ we arrive at the FE
\beq\label{1-26}
\overline{\alpha}(x,\alpha)=\bar{\alpha}(~\frac x t,~\bar{
\alpha}(t, \alpha)),\eeq
\noindent identical with eq.(\ref{1-5}).

   Note, that the corresponding continuous one-parameter transformation is
just the change ("1"~$\rightarrow $~"2") of the parameterization point
\beq\label{1-27}
R_t : \{ \mu_1 \to \mu_2 = \sqrt{t} \mu_1, ~\alpha_1
\rightarrow \alpha_2 = \albar (t, \alpha_1) \}~~. \eeq                  

  Thus, we have shown that in the renormalized QED there exists invariance
with respect to continuous transformations of the group type which involve
two quantities and contain a functional dependence. This precisely
corresponds to definition (\ref{1-10}).

 As can be shown, in QED the effective coupling $\bar{\alpha}$ is equal to a
product of $\alpha $ and dimensionless function $d(x,\alpha)$ -- the
transverse photon propagator amplitude with due regard for vacuum
polarization effects. Generally, in QFT models with one coupling constant
the invariant coupling $\bar{g}(x,g)$ can be expressed as a product of $g$,
corresponding vertex function and the square root of propagator amplitudes of
the fields participating in the interaction. Usually, this can be done on the
basis of Dyson finite renormalization transformations. \par

  Thus, the RG invariance is nothing else but the invariance of a solution
with respect to the way of its parameterization.  For instance, in real QED,
instead of using the ``Millikan's value" $~\albar (0) = 1/137~$ one may take
the ``CERN value" $\albar(M_Z^2) \simeq 1/128,9$.

\section{Renormalization group method \label{s2}}
\subsection{Basic idea \label{ss2.1}}

 Approximate solution of the physical problems with RG~ symmetry usually does
not obey this symmetry which is lost in the course of approximation.  This is
essential when the solution under consideration posseses a singularity as far
as the singularity structure commonly is destroyed by an approximation.

 In QFT, e.g., the usual way of calculation is based on the perturbation
method, i.e., on power expansion in $g$. It is not difficult to see that
finite sums of this expansion do not satisfy the functional group equations.
As the simplest illustration, consider the effective coupling $\bar{g}$ in
the UV region where the one-loop contribution has a logarithmic form
\beq\label{2-1}
\bar{g}_{\rm PTh}^{(1)}(x,g)=g+g^2 \beta \ln x~ \eeq
with $\beta$, a numerical coefficient.
 By substituting this expression into FE (\ref{1-5}), after simple
manipulation one has
$$
Discr[\bar{g}_{\rm PT}^{(1)}] \equiv \bar{g}_{\rm PT}^{(1)}(x,g) -
\bar{g}_{\rm PT}^{(1)}\left(x/t, \bar{g}_{\rm PT}^{(1)}(t,g)\right)
= $$
$$ =[g + g^2\beta \ln x]-[g+g^2\beta\ln x+2g^3\beta^2\ln t\ln(x/t)]\neq 0 $$
--- error in the $g^{3}$ order. This discrepancy can be liquidated by
addition of the particular next order term to the r.h.s. of  (\ref{2-1})
$$
\bar{g}_{\rm PT}^{(2)}= g+g^2 \beta\ln x + g^3\beta^2 \ln^2 x .$$

This ``improved" expression would yield the discrepancy of the $g^4$ order
which in its turn can be abolished by adding the $g^4\ln^3$ term to
(\ref{2-1}) and so on.

 Thus, we see, on the one hand, that the finite polynomials cannot satisfy
the condition of renormalization invariance. On the other hand, we can
conclude that the functional RG equation represents a tool for iterative
reconstruction of renorm-invariant expression that has the form of infinite
series.

  This example illustrates a rather general situation. As a rule, approximate
solutions do not satisfy a group symmetry. Here, this happens in the UV limit
as $\ln x\to \infty$ where the observed discrepancy becomes important.

  Another illustration is provided by the one-dimensional transfer problem
(for detail, see our reviews in Refs.~\cite{kiev84}). Take a half-space
($l>0$) filled with a homogenious media. Let some given amount of particles
(or radiation) be falling on the surface (at $l=0$) from the empty
half-space. The particle density $n(l,{\bf v})$ is a function of coordinate
and of particle velocity ${\bf v}$. It satisfies an integro--differential
kinetic Boltzmann equation. In some cases one can neglect by the energy
dependence of cross--sections. Here, the solution can be treated as a
function of coordinate and direction of particle velocity ${\bf \Omega} ={\bf
v}/v$. In this, ``one-velocity", case the simple symmetry of the RG type was
found not for density, but for $n$ integrated over directions in forward
hemisphere
$$
G(l) = \int_{\Omega_+} n(l, {\bf\Omega}) d{\bf\Omega}~. $$
 The function $G$ relates to amount of all particles moving inwards the
media. A partial solution of this problem will depend on the boundary
condition  at $l=0$. Correspondingly, the solution characteristic $G$ will be
the function of two arguments $G(l,g)$ -- coordinate $l$ (distance from the
boundary) and total amount of ingoing particles $g=G(l=0)=G(0,g)$. Just this
function $G(l,g)$ satisfies \cite{mamik} the group FE (\ref{1-7}) in the
additive form.

 It is rather simple to get an approximate behavior of this density $G(l,g)$,
at small $l$
\beq\label{2-2}
G(l,g) = g + l G'(0,g) ,~~~~~~~~\qquad l \ll  1 ,\eeq
which, being considered
for large $l$ values, also does not obey the mentioned symmetry.   \par

 On this basis one can set the task of ``renormalization-invariant
improvement" of  perturbative  results.  The key idea is to combine an
approximate solution with the group equations. The simplest and most
convenient way for this ``marriage" is the use of Lie equations, i.e., group
differential equations. The renormalization group method (RGM) as it was
first formulated in Refs.\cite{bs-55a,bs-55b,nc-56} is essentially
based on these group equations.

\subsection{Differential formulation  \label{ss2.2}}

  The differential equations can be obtained from the functional ones in
two different ways. Differentiating eq.(\ref{1-13}) by $x$ and putting then
$t=x\,$, one obtains (compare with (\ref{1-4})):
$$
 x {\partial \bar{g}(x,y,g)\over \partial x} = \beta \left(\frac
{y}{x}, \bar{g}(x,y;g)\right)~ ,~~~\beta(y,g)=\left.{\partial \bar{g}(t,y;g)
\over\partial t}\right|_{t=1}~. \eqno{(22a)} $$                                %
  The nonlinear equation (22a) can be considered as ``massive"
generalization of eq.(\ref{1-4}). On the other hand, one can
differentiate eq.(\ref{1-13}) with respect to $t$ at the point $t =1,$
which yields
\beq\label{2-5}
 \left[ x {\partial \over \partial x} + y{\partial \over \partial y} -
\beta (y,g){\partial \over \partial g} \right] \bar{g}(x,y;g) = 0~,\eeq
a linear partial differential equation (PDE).

   Analogous operations applied to the second of eqs.(\ref{feqs}) lead to:
$$  {\partial s(x,y;g)\over \partial \hbox{ lnx}} = \gamma        
\left[\frac{y}{x}, g(x,y;g)\right]s(x,y;g) \eqno{(22b)}  $$    
\noindent and
\beq\label{2-7}
\left\{x {\partial \over \partial x} + y {\partial \over \partial y} -
\beta(y,g){\partial\over\partial g}-\gamma(y,g)\right\}s(x,y;g)=0
\eeq
where
\beq\label{2-8}
\gamma(y,g) =\left.{\partial s(t,y;g\over \partial t}\right|_{t=1}\eeq
is the so-called {\it anomalous dimension} of $s$. For a group invariant,
like, e.g., matrix element $M$ satisfying FE (\ref{1-18}) this dimension is
equal to zero. The corresponding PDE looks like
\beq\label{2-9}
\left\{\sum^{}_i x_i {\partial\over\partial x}_i-y{\partial\over\partial y}
-\beta(y,g){\partial\over\partial g} \right\} M(x,y;g)=0 ~.\eeq

 Equations (\ref{2-5}), (\ref{2-7}), and (\ref{2-9}) express the independence
on the $t$ parameter of the r.h.s. of the related functional group equations,
i.e., a mutual compensation of $t$ dependences via three (or more) arguments.
This DEs can be called {\it compensational} equations to distinguish them
from nonlinear eqs.(\ref{1-22}) which can be referred to as {\it evolutional}
group equations.

 Stress that compensational as well as evolutional DEs taken together with
normalization (i.e. boundary) conditions like $\bar g(1,g)=g~,~s(1, g)=1 $
are equivalent to functional equations and to each other. At the same
time, evolutional Lie equations turn out to be more convenient for practical
construction of the solution, generators $\beta, \gamma$ being given.

   Let us comment also that the UV limit of compensational DEs like, e.g.,
\begin{equation}\label{2-11}
\left\{x {\partial \over \partial x} -
\beta (g){\partial \over \partial g} - \gamma (g)\right\} s(x,g) = 0
\end{equation}
 coincides with the UV limit of specific nonclosed equation
$$
\left\{x {\partial \over \partial x} - \beta (g){\partial
\over \partial g} - \gamma (g)\right\} s(x,g) = \Delta  S \eqno(C-S) $$
obtained in the early 70s by Callan and Symansik.  The r.h.s. of
this equation contains the result of mass counter-term insertion into all
internal lines of all diagrams for the function $s$ under consideration. For
this reason, in current literature compensational equations are often related
to as the Callan--Symansik equations. However, these equations just in the
form (\ref{2-5}) and (\ref{2-7}) were first obtained by Lev Ovsyannikov in
1956 while solving \cite{oves56} functional RG equations. Therefore, we
consider it justifiable to relate compensational DEs to the Ovsyannikov's
rather than to some other names. \par

   It is not difficult to formulate group DEs for a multi-coupling case by
proper differentiation of FEs (\ref{1-16}). For instance, the system of
evolutional DEs looks like
\beq\label{2-10}
x{\partial \bar{g}_i(x,y,\{g\})\over \partial x} =
\beta_i\left(\frac{y}{x}, \{\bar{g_j}(x,y;\{g\})\}\right)~. \eeq      

\subsection{General solution \label{ss2.3}}                      

 General solution of the group FEs was obtained in the paper \cite{oves56} by
 applying the theory of PDE to the compensational eqs. (\ref{2-5}) and
(\ref{2-7}). Details of the derivation can be found in the Section 48.3 of
the third edition of the monograph \cite{Book}. The results obtained can be
formulated as follows:  \par

 To every solution of DE (\ref{2-5}), there corresponds some function of two
arguments $F(y, g)$, reversible with respect to its second argument and
connected to $\bar{g}$ by the relation
\beq\label{2-12}
 F(y,g) = F\left(\frac{y}{x},\; \bar{g}(x,y;g)\right)~. \eeq   
The explicit form of $\bar{g}$
can be obtained now by reversing the r.h.s:
 $$
\gbar (x, y, g) = F_{(2)}^{-1} \left[\frac{y}{x}, \;F(y, g)\right]~. $$
To determine $F$ it is sufficient to specify the generator $\beta (y,g)$.

  Note also that to get from the Ovsyannikov result (\ref{2-12}) the solution
in the UV limit, i.e., in a massless case at $y = 0,$ one has to assume for
$F$ a specific limiting form
$$
F(y,g)=y \exp [f(g)]~~~\hbox{or}~~= \ln y+f(g)~~~\hbox{as}~~~~y \to 0~.$$
Then
\beq\label{2-13}
f\{\bar{g}(x,g)\} - f(g)=\ln x~;~~~
\gbar = f^{-1} \{ \ln x + f(g) \} ~. \eeq                   

 Here, $f'(g) = 1/\beta (g)$. This is equivalent to the Gell-Mann---Low---Lee
solution (\ref{1-19}).

  To every solution of eq.(\ref{feqs}) for a function $s$,
there corresponds  some function $\Sigma (y,g)$ related to $s$ by
\beq\label{2-14}
s(x,y;g) = {\Sigma [y/x,\bar{g}(x,y;g)]\over \Sigma (y,g)}~.\eeq     
Let us give also the general solution of the same type for the system
(\ref{1-16}) for the $k$-couplings
case.  It can be written down in terms of $k$ arbitrary functions
$F_i$, reversible simultaneously with  respect to last
arguments, and defined from the system of $k$ functional relations
\beq\label{2-15}
F_{i}(y,\{g\}) = F_i\left[\frac{y}{x}~,~\{\bar{g}(x,y;\{g\})\}\right]~,~~
 \{g\} = g_1,...~g_k~~;~ i,j = 1, ...~k~. \eeq

 All solutions (\ref{2-12}) --- (\ref{2-15}) satisfy the usual normalization
conditions.

   The transition to the massless limit in expressions (\ref{2-14}) ---
(\ref{2-15}) can be performed by a trick analogous to the given above.
Then, e.g.,
$$ 
s(x,g)=x^{\gamma}\frac{\Sigma\{\bar{g}(x,g)\}}{\Sigma (g)}~. $$

Let us also formulate solution for the 2-coupling case $\,g_1=g,~g_2=h\,$
in the massless limit in the form analogous to (\ref{2-13})
\beq\label{2-17}
f_{i}(\bar{g},\bar{h})=f_{i}(g,h)+\ln x,~~~~~ i=1,2~.\eeq

From the solutions presented it follows that imposing group symmetry one
reduces by unity the number of independent arguments.

\subsection{RGM algorithm  \label{ss2.4}}
\subsubsection{\small\it Technology of RG Method \label{sss2.4.1}}

  The idea of the approximate solution marriage~\cite{bs-55a,bs-55b} with
group symmetry can be realised with help of group DEs. If we define group
generators $\,\beta, \gamma\,$ from some approximate solutions and then
solve evolutional DEs, we obtain {\it RG improved} solutions that obey the
group symmetry and correspond to the approximate solutions used as an input.
\smallskip

   Now we can formulate an algorithm of improving an approximate solution.
The procedure is given by the following recipe which we illustrate
by a massless one--coupling case (\ref{1-4}) and (\ref{1-5}):
\par
Assume some approximate solution $\,\gbar_{\rm appr}\,$ is known.

{\bf 1.} On the basis of eq.(22a) define the beta-function
\beq\label{2-18}
\beta (g) \define {\partial\over{\partial \xi}} \gbar_{\rm
appr} (\xi, g) \bigg|_{\xi = 1} ~~.  \eeq                   

{\bf 2.} Integrate eq.(\ref{1-4}), i.e., construct the function
\beq\label{2-19}
f(g) \define\int\nolimits^g{{d\gamma}\over{\beta (\gamma)}}~,\eeq 

{\bf 3.} Resolve the eq.(\ref{2-13})
\beq\label{2-20}
 \gbar_{\rm RG} (x, g)= f^{-1} \{ f(g) + \ln x \} ~~. \eeq

{\bf 4.} Then, the solution $\gbar_{\rm RG}$, precisely satisfies the RG
symmetry, i.e., it is an exact solution of eq.(\ref{1-5}) and corresponds to
$\,\gbar_{\rm appr}\,$.
\smallskip

For illustration, take as a $\gbar_{\rm appr}$ the simplest perturbative
expression (\ref{2-1}) for the invariant coupling. Here, the $\beta$-function
is $\beta(g)=-\beta_1 g^2~,$ and the integration yields
$$
\int_g^{\gbar}{{dg}\over{\beta (g)}} = {1\over \beta_1}\left(
{1\over\gbar} - {1\over g}\right) = \ln x ~.$$                    

The solution obtained
\begin{equation}\label{2-21}
\gbar(x,g)={g\over 1+g \beta_1\ln x }~ , \end{equation}
one one hand, exactly satisfies the RG symmetry and, on the other, being
expanded in powers of $g$, correlates with the input (\ref{2-1}).

\subsubsection{\small\it RGM usage in QFT \label{ss2.5}}

As it has been explained above in Section \ref{ss2.1}, the QFT perturbation
expression of finite order does not obey the RG symmetry. On the other hand,
in Section \ref{sss2.4.1} it was shown that the one-loop UV approximation for
$\gbar$ used as an input in eq.(\ref{1-23}) for the construction of a group
generator $\beta(g)$ yields expression (\ref{2-21}) that obeys the group
symmetry and exactly satisfies FE (\ref{1-5}).

Now, using the geometric progression (\ref{2-21}) as a hint, let us represent
the 2-loop perturbative approximation for $\,\gbar\,$ in the form
$$
 \gbar_{\rm pt} = g - g^2 \beta_1 \ln x + g^3 ~[ \beta_1^2
\ln^2 x - \beta_2 \ln x ] + O(g^4) ~~, $$
where $\beta_1\,$ and $\beta_2\,$ mean the $\beta\,$-function coefficients
at the one-loop and two-loop level, respectively.
If we substitute this expression into eq.(\ref{1-5}) we obtain
$$
\gbar^{(2)}_{pt}(x,g)- \gbar^{(2)}_{pt}(x/t;\gbar^{(2)}_{pt}(t,g))=
g^4\beta^3_1 \ln (x/t) \ln^2 t~.$$

Meanwhile, we can use $\gbar^{(2)}_{pt}$ as an input in Eq.(\ref{2-18}). Now
 the step {\bf 1} yields
$$ \beta^{(2)}(g)= -\beta_1 g^2 -\beta_2 g^3 $$
and then (step {\bf 2})
\beq\label{2-22}
\beta_1f^{(2)}(z)= - \int^z{d\gamma\over\gamma^2+b\gamma^3}={1\over z}+
b \ln {z \over 1+bz}~;~~~b={ \beta_2 \over \beta_1}~.\eeq  

To make the last step, we have to start with the equation
$$
f^{(2)}[\gbar^{(2)}_{\rm rg}(x,g)]=f^{(2)}(g)+ \beta_1\ln x $$
which is a transcendental one and has no simple explicit solution\footnote{It
can be expressed in terms of a special, Lambert, $W$-function :
$W(z)\exp^{W(z)} =z$; see, e.g., \cite{tmf99}.}. Due to this, one has to
resolve this relation approximately. Take into account that the second,
logarithmic, contribution to $f^{(2)}(z)$ in (\ref{2-22}) is a small
correction to the first one at $bz\ll 1$. Under this reservation we can
substitute the one-loop RG expression (\ref{2-21}) instead of
$\gbar^{(2)}_{\rm rg}$ into this correction and obtain the explicit
expression
\beq\label{2-23}
\gbar_{\rm rg}^{(2)} = {g\over{1 + g\beta_1 l + g(\beta_2/\beta_1)
 \ln ~[ 1 + g \beta_1 l ] }} ~;~~~~ l = \ln x. \eeq               

This result (first obtained \cite{bs-55b} in mid-50s) is interesting in
several aspects.

First, being expanded in $g$ and $gl$ powers, it produces an infinite
series containing ``leading", i.e. $\sim g(gl)^n$, and ``next-to-leading"
$\sim g^2(gl)^{n}$ UV logarithmic contributions.  Second, it contains a
nontrivial analytic dependence
 $$
\ln (1+g\beta_1l) \sim \ln(\ln Q^2)$$
 which is absent the in perturbation input.
Third, being compared with eq.(\ref{2-21}), it demonstrates algorithm of
subsequent improving of accuracy, \i.e., of RGM regularity.
\medskip

  Now we can resume the RGM properties. The RGM is a regular procedure of
{\it combining} dynamical information (taken from an approximate solution)
with the RG symmetry. The essence of RGM is the following:
\smallskip

\noindent
1) The mathematical tool used in RGM is Lie differential equations.
\smallskip

\noindent
2) The key element of RGM is possibility of (approximate) determination of
group generators from dynamics.
\smallskip

\noindent
3) The RGM works effectively in the case when a solution has a singular
behaviour.  It restores the structure of singularity compatible with RG
symmetry.

\section{RG in QFT \label{s3}}                     

This section is devoted, mainly, to general topics of RG applications in the
QFT short--distance asymptotic behavior. We discuss the specific features of
UV analysis connected with use of perturbation theory, in particular,
reliability of results.

\subsection{UV analysis in general \label{ss3.1}}
 \subsubsection{\small\it One-coupling case  \label{ss3.1.1}}

General analysis of the UV asymptotic behavior for the one-coupling QFT model
can be performed rather simply on the basis of the solution
\begin{equation}\label{3-1}
\int\nolimits^{\bar g(x,g)}_g {d\gamma\over \beta(\gamma)} = \ln x~,~~
x ={Q^2\over \mu^2}~~, \end{equation}
of the massless RG equation (\ref{1-4}) for an effective coupling with
$g=\bar g(1, g)$ and $\mu$, a reference point. As follows from it, the
asymptotics at $\ln x \to\infty$ corresponds to the divergence at the upper
limit of the l.h.s. integral. Depending on the feature of the
$\beta$--function the resultant UV behaviour of the invariant coupling $\bar
g$ differs very much.

 Suppose that at very small $g$ values the beta-function is positive.
 Then, three cases are possible :
 \medskip
\par ${\bf a)}$ Consider first the situation with
$$
\int_g^{g_{*}}{dz\over \beta (z)} = \infty \eqno(52a)$$                         
corresponding to the case when the beta--function has a zero at some finite
point $\,g_*\,$.

 Here, the UV asymptotic value of effective coupling is finite
$\bar{g}(\infty ,g)=g_{*} < \infty~,$
which relates to the {\it finite renormalization of the coupling constant}:
$Z=\bar{g}(\infty,g)/g $.

Using the terminology of DEs qualitative theory, one can say that at
$g=g_*\,$ we have  a UV fixed point.

 If at $g=g_{*}$ there is a first order zero
$\beta(g)\simeq b(g_*-g)~,$ then eq.(52a) gives
 \addtocounter{equation}{1}
\beq\label{3-3}
 \bar{g}(x,g)-g_{*} \simeq C \exp^{-b\ln x}=C(Q^2/\mu^2)^{-b}~~\hbox{as}~~~
Q^2 \to \infty~, \eeq
i.e., in the vicinity of a fixed point we have an asymptotic {\it power}
regime.  \medskip

\par {\bf b)} If $\beta(g)$ is (monotonically) increasing as $g \to \infty$
but gentler than $g^2$, so that
$$
\int_g^{\infty}{dz\over \beta (z)} = \infty \eqno(52b)$$
then the effective coupling tends to infinity
$$
\lim_{x\rightarrow \infty} \bar{g}(x,g)\to \infty~, $$
which corresponds to infinite coupling constant renormalization.  Formally,
this is equivalent to $g_{\infty }=\infty$.  \medskip
\par ${\bf c)}$ At
 $$
 \int\nolimits^\infty_g{d\gamma\over \beta(\gamma)}
= L =\ln x_{\infty}<\infty ~,\eqno(52c)$$
that happens if
  $$
\lim_{x\rightarrow\infty}\beta(g)/g^2 \ge\hbox{const} ~,$$       
the theory has an inner contradiction, as far as
 $$
\bar{g}(x_{\infty},g)=\infty~~~\hbox{at}~~~x_{\infty }<\infty $$
and the momentum region $x > x_{\infty }$ can not be described by the theory.
 We encounter here a {\it ghost trouble}, as explained below in this Section.

Up to now we have assumed that the generator $\beta (g)$ is positive.
In the opposite case
\medskip

{\bf d)} $\beta(g)=-b(g)< 0\,\,$ one has to deal with the equation
$$
\int^g_{\bar{g}}{dz\over b(z)}=\ln x \eqno(52d)$$                
and study possible divergence of the integral involved at the lower limit.

If this occurs at some finite value $g = g_{\infty}\,$, the situation is
quite analogous to the case {\bf b)}.  The only difference is that now the
effective coupling tends to its limiting value $g_{\infty }$ from above.

 As the most important case, we consider the possibility when the singularity
lies at the origin $\beta(0)=0\,$ which happens in QCD. Then, $\bar{g}$
vanishes $\bar{g}(\infty, g) = 0\,$ in the UV limit which corresponds to the
{\it asymptotic freedom} phenomenon. E.g., if we assume here that~~
$\beta(g)=-\beta_1g^2$ at $g\to 0\,,$ then
\beq\label{3-4}
\bar{g}(x,g) \to {1 \over \beta_1 \ln x}~~~~\hbox{as}~~~~x \to \infty~.\eeq

 \subsubsection{\small\it Multi-coupling case  \label{ss3.1.2}}

 For the quantum field model with several coupling constants one has to
consider the system of coupled functional (\ref{1-16}) or differential
equations. The last ones can be analyzed by the well-known methods of the
qualitative theory of differential equations.

 Take the case with two coupling constants $g$ and $h$.
The system of evolutional differential equations is
\beq\label{3-5}
\dot{\bar g}=\beta_g (\bar{g}, \bar{h})~,~~\dot{\bar{h}}=\beta_h (\bar{g},
\bar{h})~;~~\dot{f}\equiv \partial f/\partial \ell~;~~\ell=\ln x~. \eeq

According to (\ref{2-17}), the general solution to this system is of the form
$$
F(\bar{g}, \bar{h}) = F(g, h) + \ell~,~~~~ \Phi (\bar{g}, \bar{h})
= \Phi (g, h) + \ell~.$$
where $F$  and $\Phi$ --- two arbitrary reversible functions.

As far as argument $l = \ln x$ does not enter explicitly into the
generators $\beta_g $ and $\beta_h$, it can formally be excluded by dividing
one of the equations (\ref{3-5}) by the other :
\beq\label{g-h}
{d\bar{g}\over d\bar{h}}=F(\bar{g},\bar{h}),~~~~F={\beta_g\over\beta_h}~.\eeq
This equation can be analyzed on the two dimensional phase plane
$(\bar g,\bar h)$.

First explicit example of such phase portrait has been obtained in
mid-fifties by I. Ginzburg \cite{ilya} --- see also Section 51.4 in the
third edition of the monograph \cite{Book}.                       
 The essential features are now singular points and singular solutions.
 Singular points correspond to $\beta_i=0$ (or $=\infty$). They can be of
different types: a stable fixed point that is known as {\it attractor}, an
unstable fixed point and a saddle-type point. In the vicinity of the UV
attractor one can have a power scaling behavior as in eq.(\ref{3-3}).
Singular solution, separatrix, joins singular points and can also be stable
or unstable. Generally, the unstable ones separate the parts of phase plane
with different UV asymptotes that correspond to UV stable separatrices.
\vskip 0.5cm

\subsection{Perturbative approach to the UV asymptote \label{ss3.2}}
\subsubsection{\small\it Structure of RG results  \label{sss3.2.1}}

  Consider a general situation with the RG approach to the UV asymptotic
behavior based on perturbation calculation input. In the one-coupling QFT
case, group generators entering into DEs can be written as
\beq\label{3-6}
\beta (g) = \beta _{1}g^2 + \beta _{2}g^3 + \ldots ~~,~~~~~~
 \psi (g) = \psi _{1}g + \psi _{2}g^2 + \ldots   .\eeq
Generally, expansion coefficients depend on the mass variable
\beq\label{3-7}
\beta (y,g) = \sum^{}_{l\ge 1} \beta _{1}(y)g^{l+1}~~, ~~~~~~~~
\psi(y,g) =\sum^{}_{l} \psi _{l}(y)g^{l}~ .\eeq     %

Note that if $g$ is just the $S$--matrix expansion parameter (that can be
equal to the coupling constant or to its square) then usually the first term
in expansion for $\beta $ is quadratic and for $\psi $ --- linear, as it is
explicitly indicated above.\par

Substituting (\ref{3-6}) into (\ref{3-1}), and re-expanding the ratio
$\gamma^2/\beta(\gamma)$, we obtain after integration
\beq\label{3-8}
1/g - 1/\bar{g} - {\beta_1 \over \beta_2 } \ln  (\bar{g}/g) -
b_{3}(\bar{g} - g) + 0(\bar{g}^2,g^2) = \beta _1\ln x~ ,\eeq
\beq\label{3-9}
\ln s(x,g)=(\psi_1/\beta_1)\ln\{\bar{g}(x,g)/g\}+c_2(\bar{g}-g)+0(g^2)~,\eeq
$$
b_{3} = \beta _{3}/\beta _{1} - (\beta _{2}/\beta _{1})^{2} ,~~~ c_{2}
= [\psi _{2}/\beta _{1}](\psi _{2}/\psi _{1} - \beta _{2}/\beta _{1}) .$$

As follows, the solutions $\bar{g}$ and $s$ depend on two arguments $g$ and
$g \ln x$ . By expanding them in powers of $g$ we get
\beq\label{3-10}
\bar{g}(x,g) = g f_{1}(g \ln  x) + g^{2}f_{2}(g \ln  x ) + \ldots~,~~~
\ln s(x,g) = \varphi _{1}(g\ln x)+g\varphi_{2}(g \ln x) +\ldots~ \eeq
where $f_{j}$ and $\varphi _{i}$ have a simple form. For example,
$f_1(z)=(1-\beta_1 z)^{-1}$, $\varphi_1(z)\sim f_2(z)\sim\ln f_{1}(z)~.$

Comparing expressions obtained with usual perturbative expansions
$$
\bar{g}_{pt}(x,g) = g + g^2\beta _{1}\ln  x + g^{3}[
\beta ^{2}_{1}\ln  x + \beta _{2}\ln  x ] + 0(g^{4}) , $$
$$
s_{pt}(x,g) = 1 + g\psi _{1}\ln  x + g^{2}[(\beta _{1}
\psi _{1}/2)\ln ^{2}x + \psi _{2}\ln x] + 0(g^{3})~, $$
used as an input to obtain our starting generators, one can see the
qualitative effect of the RGM using. In the case considered, it changes the
region of applicability of the perturbation method limited by the condition
$g\ln x \ll  1$ to a more larger region defined by two relations
\beq\label{3-12}
g \ll  1~ ,~~~~~~~~ \bar{g}(x,g) \ll  1 , \eeq
the second of which is defining.

\subsubsection{\small\it The ghost-pole trouble \label{sss3.2.2}}

Turn now to the one-loop RG approximation for the effective coupling
$\bar{g}$, considered in the UV, i.e., massless limit.
 According to (\ref{2-21}) and (\ref{a1rg}), it has the form
$$
\bar{g}_{(1)}(x,g)={g\over 1-\beta_{1}g \ln x}~.$$       
Let the numerical coefficient $\beta_1$ be positive. Such is the case in QED
where $\beta _1= 1/3\pi$ and $g$ stands for the expansion parameter
$\alpha=e^2 $. This expression obviously has a pole singularity at
$$
x = x_{*} \equiv \exp (1/\beta _{1}g) = \exp ( 3\pi /\alpha )~.$$     

As far as the QED effective coupling is proportional to the (transverse part
of) photon propagator, this pole, generally, describes some bound state
of a system with the photon quantum numbers. However, a pole related to a
physical bound state must have positive residue while the l.h.s. of
eq.(\ref{a1rg}) has a negative one.

 This means that it corresponds not to a physical but rather to some
unphysical, so-called {\it ghost\/}, state. The presence of a ghost
singularity can be treated as a signal of inconsistency of a theory. Such
claims have been made~\cite{zero-trouble} in the mid 50s when the ghost-pole
trouble was first discovered~\cite{zero} just before the birth of the RGM.

  The RG method proved to be very effective for a general discussion of
the ghost-pole issue. The first question that must be answered here is the
stability of indication of the ghost-pole existence with respect to the
multi-loop corrections.
\medskip

Note that in a perturbation calculation, the $\beta$--function depends on the
adopted renormalization procedure; at the massless case, starting with the
3-loop level the coefficients of the perturbation series (\ref{3-6}) depend
on the renormalization scheme (RS) used. In QED, the 3-loop $\beta$ function
in MOM (i.e., momentum subtraction) scheme is
\begin{equation}\label{3loop}
\beta^{MOM}_{(3)}={\alpha^2\over 3\pi}+{\alpha^3\over 4\pi^2}+{\alpha^4\over
8\pi^3}\biggl( {8\over 3}\zeta(3)-{101\over 36} \biggr)~.  \end{equation}

The numerical value of the last parenthesis is about $0.4\/$. Neglecting
it for the moment, we start our discussion with the two-loop approximation
for the $\beta\/$ function. According to eq.(\ref{a2rg}), the 2-loop
iterative RG solution is
$$
\bar\alpha_{(2)}={\alpha\over 1-{\alpha\over 3\pi} l +{3\alpha
\over 4\pi}\ln(1-{\alpha\over 3\pi} l )}~~. $$            %

 This solution has an error of an order of $\alpha^4 l\/$ and is interesting
from several points of view. As it has been mentioned before, its $\alpha$
expansion besides leading logs contains an infinite number of next-to-leading
terms $\alpha^2(\alpha l)^m$, the first of which has been used as an input
for construction of the $\beta$ function. Second, in the vicinity of the
ghost pole of the one-loop RG solution at $l_1=3\pi/\alpha$, the two--loop
$\bar\alpha_{(2)}$ solution differs from $\bar\alpha_{(1)}$ considerably.
 Hence, an infinite sum of the next-to-leading logarithmic contributions in
the region $\alpha l\sim 1$ becomes important.

It is not trivial because for an each order of the perturbation input the
next-to-leading term is negligible comparing with the leading one of the
previous order (the ratio being of an order of $\alpha\beta_2 / \beta_1 =
3\alpha/4\pi \approx 2.10^{-3})$.  It can be seen with the help of
(\ref{3-8}) that the allowing for the last, 3-loop, term in (\ref{3loop})
also becomes essential for the $\bar{\alpha}\sim 1$ case.

 This means that the problem of existence for the ghost pole in QED cannot be
solved by taking into the account of next-to-leading and so on logs.
Moreover, one can argue~\cite{bsh56} on general RG ground that it is
impossible to make any qualitative statement about the UV asymptote for the
$\beta(g)>g\/$ case basing on RG-improved perturbation calculations. Our next
example illustrates this thesis.

\subsubsection{\small\it Scalar quartic model \label{sss3.3.2}}     

For the nonlinear scalar field with the quartic (self)interaction Lagrangian
$$
{\cal L}_{int}= -{4\pi^2\over 3}g\phi^4~ $$
important progress has been achieved in 80s in the higher perturbation orders
calculation. The $\beta\/$ function was calculated in the MS-scheme up to the
5-loop level \cite{4loop}
$$ 
\beta^{MS}_{(5)}={3\over 2}g^2-{17\over6}g^3+16.275g^4-135.8g^5+1437g^6~.$$

We see from this expression that, in conrast with the QED case, due to its
alternate-sign structure, there is no stability here even on a qualitative
level. The odd-order approximations have a ghost-pole type behavior, whereas
the even ones yield the fixed point (finite charge renormalization) case.
Note also that, as can be shown~\cite{kazak}, the upper boundary of the 10\%
confidence  region corresponds to $g$ values close to 0.1 .

  To comprehend the ``loop dependence" of this boundary it is useful to
represent expression $\beta^{MS}_{(5)}$ in a slightly different form
\beq\label{beta5a}
\beta^{MS}(g)={3\over2}g^2\biggl[1-{g\over 0.529}+
\biggl({g\over0.303}\biggr)^2-\biggl({g\over 0.222}\biggr)^3+
\biggl({g\over 0.180} \biggr)^4\biggr]~~. \eeq

 It is clear now, that a boundary of the confidence region diminishes with
rising the order of a loop approximation. The expression (\ref{beta5a}) looks
like a beginning of a power asymptotic series of the Poincar\'e type. Indeed,
if we represent the $\beta\,$-function in a series expansion form (\ref{3-6})
then, as it is can be shown, the coefficients $\beta_n\,$ at $n\gg 1$ behave
like $\sim n!$.

 The method of determining asymptotic estimates of the perturbation expansion
coefficients of the Green functions uses a representation in the form of a
functional (i.e. path) integral. This integral written down for the mentioned
expansion coefficient can be calculated by the steepest descent method in the
function space. To the saddle point, there corresponds an ``instanton"--type
Euclidean classical solution with a finite action.

  In this manner, an asymptotic expression was obtained~\cite{lipat}
 for the coefficients of the $\beta\,$ function expansion. It has the form
\beq\label{lip}
\beta_n\simeq {1.096\over 16\pi^2} \/n! ~n^{7/2}\left(1+O(n^{-1})\right)
\quad\qquad (n \rightarrow \infty)~.\eeq
The factorial growth of coefficients indicates that this is a power
asymptotic series with zero radius of convergence that cannot be
summed in the usual manner. We can obtain the information about the
singularity structure at the origin ($g=0$) by using some special
procedures. One of them is the Borel summation method.

  Here, we give short exposition of the results on the attempt of the
summation of the series of (\ref{beta5a}) type made in \cite{kazak} (see also
the review \cite{kazak-sh}).

  Authors of the Ref.~\cite{kazak} used as an input the $\beta$-function
4-loop expression in the symmetric MOM-scheme
\beq\label{beta5mom}
\beta^{MOM}_5(g)={3 \over 2}g^2-{17 \over 6}g^3+19.3g^4-146g^5 ~.\eeq

This alternate-sign asymptotic series can be summed by the Borel method. The
idea is to represent the sum in a form of a Laplace transform integral. It is
not difficult to see that the transition to the Laplace image just ``kills"
the factorial factor $n!$. For the modified Borel transformation
\beq\label{borel-tr}
\beta(g)=\int\nolimits^{\infty}_0 {dx\over g}\exp{(-x/g)}
\biggl(x{\partial\over\partial x}\biggr)^5  B(x)~ \eeq
 perturbation series can be written down as
$$
B(x)=\sum_n {\beta_n\over n!n^5}x^n~~. $$
  It has a nonzero circle of convergence and can be summed within the circle.
 However, as the integration domain in (\ref{borel-tr}) goes outside the
convergency region, we must make an analytic continuation for the function
$B(x)$. It can be done by a conformal transformation of the $x$-plane into
the $w$-plane to map the domain of integration $[0,\infty]$ into the interior
of the unit disk and the cut $[-\infty,-1]$ into the boundary of the disk.
One can choose this transformation in such a way that it correctly reproduces
the singularity on the cut. The result of conformal transformation $x\to
w(x)=(\sqrt{1+x}-1)\cdot(\sqrt{1+x}+1)^{-1}$  ``looks quite well" :
$$
B(x)={3x^2\over 128}(1-0.32w-0.127w^2+0.084w^3)~.$$

 Then, by transformation reverse to (\ref{borel-tr}) one can reconstruct beta
function $\tilde{\beta}(g)$ which is nonanalytic in the $g$ variable with
essential singularity at $g=0$. Graphs of the function $\tilde{\beta}(g)$
obtained by the Borel summation with allowance for the 1-, 2-, 3- and 4-loop
approximations look very similar one to other.

  They all lie now in a narrow parabolic ray slightly below the original
one-loop parabola and within the limit of 10\% accuracy enable to advance
into the region $g\sim 50$. This means that the summation procedure adopted
enlarges the confidence interval in several hundred times! Besides this it
gives the qualitative stability of results. All they are now in favour of a
ghost-type UV asymptote.

  Nevertheless, these results can be considered only as a support but not the
proof of the $\phi^4$-model inconsistency. The weak point here is that
starting with (\ref{borel-tr}) we have assumed the definite analyticity
properties of $\beta(g)$ in the whole complex $g$-plane.

\subsection{Mass--dependent analytic solution \label{ss3.4}}

  A general method of an approximate solution to the {\it massive} (i.e.,
mass-dependent) RG equations was developed in Ref.\cite{dv81}. Analytic
expressions of a high level of accuracy for an effective coupling and a
one-argument function were obtained up to 3-- and 4--loop order \cite{92-3}.

   For example, the two-loop massive RG--solution for the invariant coupling
\begin{eqnarray}          \label{m2loop}
\alpha_s(Q^2)_{\rm rg}^{(2)} = \frac{\alpha_s}{1+\alpha_s A_1(Q^2, m^2)+
\alpha_s A_2/A_1)\,\ln \left(1+ \alpha_s A_1(...) \right)}\,
\label{a2rgm}\end{eqnarray}
at small $\,\alpha_s$ values corresponds to perturbation expansion
$$
\alpha_s(Q^2)_{\rm pert}^{(2)}=\alpha_s-\alpha_s^2 A_1(Q^2, m^2)    
+ \alpha_s^3\left[A_1^2-\,A_2(Q^2, m^2)\right] + \dots \, \,. $$
At the same time, it smoothly interpolates between two massless limits
(with $A_{\ell} \simeq \beta_{\ell} \ln Q^2 +c_{\ell}$) at $\,Q^2 \ll m^2$
and $Q^2\gg m^2\,$ described by an equation analogous to (\ref{2-23}). In
the latter case it can be represented in the form usual for the QCD practice:
$$
\bar{\alpha}^{-1}_s(Q^2/\Lambda^2)_{\rm rg}^{(2)}\simeq\beta_1\left\{\ln
\frac{Q^2}{\Lambda^2}+b_1\ln\left(\ln\frac{Q^2}{\Lambda^2}\right)\right\}\,;
\;\; b_1= \frac{\beta_2}{\beta_1^2} \,. $$

 Solution (\ref{a2rgm}) demonstrates, in particular, that the threshold
crossing generally chan\-ges the subtraction scheme \cite{mass95}.

  The investigation \cite{dv81,92-3} was prompted by the problem of taking
explicitly into account of heavy quark masses in QCD. However, the results
obtained are important from a more general point of view for a discussion of
the scheme dependence problem in QFT. The method used could also be of
interest for RG applications in other fields within the situation with
disturbed homogeneity, such as, e.g., intermediate asymptotics in
hydrodynamics, finite-size scaling in critical phenomena and the excluded
volume problem in polymer theory.  \medskip

 In paper \cite{92-4}, this method was applied to the evolution of effective
gauge couplings in Standard Model (SM). Here, a new analytic solution of a
coupled system of three mass-dependent two-loop RG equations for three SM
gauge couplings was obtained.

 For this goal, one has to start with a perturbative input for the SM
couplings
\beq\label{mass1}
\alpha_i(Q^2, m^2)\simeq \alpha_i-\alpha_i^2A_i(Q, m, \mu)+\alpha_i^3
A_i^2(Q)-\alpha_i^2\sum_j \alpha_jA_{ij}(Q)~;~~~i=1,2,3.  \eeq  
where $A_i$ and $A_{ij}$ are one- and two-loop mass- and $\mu$-dependent
contributions of appropriate Feynman diagrams. In the framework of a
massive renorm-group formalism \cite{bs-55a,bs-55b,nc-56} the corresponding
Lie equations look like
\beq \label{mass2}
\dot{\alpha}_i(Q^2,m^2)\simeq -\alpha_i^2(Q)\left[ \dot{A}_i(Q)+
\sum_j \alpha_j(Q)\dot{A}_{ij}(Q) \right]    \eeq
with $\dot{A}\equiv \partial A/\partial \ell ;~ \ell=\ln Q^2/\mu^2~$. Note
that in the UV limit $ A_i(Q)=\beta_i\ell~;~~A_{ij}(Q)=\beta_{ij}\ell~ $ we
arrive at the system
$$
\dot{\alpha}_i(\ell)=-\left(\beta_i+\sum_j \beta_{ij}\alpha_j(\ell)
\right)\alpha^2_i(\ell) $$
that is commonly used -- see, e.g., Refs.\cite{Ugo1} -- for the discussion
of data extrapolation across the gauge desert and possibility of Grand
Unification.  \par

The latter system can be solved iteratively in the form
\beq \label{sm2loop}
\frac{1}{\alpha_i(\ell)}=\frac{1}{\alpha_i}+\beta_i \ell+\sum_j \frac{
\beta_{ij}}{\beta_j}\ln[1+\alpha_j\beta_j\ell] ~;~~~ \alpha_i=\alpha_i(\mu)~.
\eeq     

Here, we present a generalization of this solution for the massive case, that
is convenient for taking into account of threshold effects and discussing, in
particular, the issue of the Grand Unification consistency check.

  Using the method of the paper \cite{Shi81}, one can obtain explicit
iterative solution to the system (\ref{mass2}). Here, as in the massless
case, one first solves the one-loop approximation to (\ref{mass2}) to get
\footnote{This exact solution of an one-loop massive RG equation was first
obtained in Ref.\cite{Blank}.}
$$
\alpha_i^{(1)}(Q^2, m^2)=[1/\alpha_i+A_i(Q^2, m^2)]^{-1}~. $$
Inserting then this explicit expression into the second factor in the r.h.s.
of (\ref{mass2}) and performing an approximate integration of some integral
--- for detail see paper \cite{Shi81} --- we arrive at the expression
\beq\label{sm2mass}
 \frac{1}{\alpha_i(Q^2,m^2)}= \frac{1}{\alpha_i}+A_i(Q)+
\sum_j \frac{A_{ij}(Q)}{A_j(Q)}\ln[1+\alpha_jA_j(Q)]~. \eeq   
quite analogous to (\ref{m2loop}).

The remarkable feature of this solution is that it depends explicitly only on
mass-dependent perturbation coefficients $A_i(Q),~A_{ij}(Q)$ and, being
expanded in powers of coupling constants, exactly corresponds to the
perturbative input (\ref{mass1}). On the other hand, in the massless limit it
goes to solution (\ref{sm2loop}).  \par

 The accuracy of the last approximate expression can be estimated by the
method used in paper \cite{Shi92}. Generally, it corresponds to the accuracy
\footnote{See eqs.(13) and (16) in Ref.\cite{Shi92}.} of three--loop
expression $(\,\simeq\alpha^5\ln \,)$ for the effective coupling in the
one-coupling case that is quite sufficient for current pactice.

\subsection{Some important results  \label{ss3.5}}              

In the early 70s, S. Weinberg \cite{steve} proposed the notion of a
 {\sl running mass} of a fermion. If considered from the viewpoint of
paper \cite{tolia}, this idea can be formulated as follows:

{\sl any parameter of the Lagrangian can be treated as a (generalized)
coupling constant, and its effective counterpart should be included into the
renorm-group formalism}.
\medskip

  New possibilities for applying the RG method were discovered when the
technique of operator expansion at short distances (on the light cone)
appeared \cite{conus}. The plausibility of this approach stems from the fact
that the RG transformation, regarded as a Dyson transformation of the
renormalized vertex function, involves the simultaneous scaling of all its
invariant arguments -- normally, the squares of the momenta. Meanwhile, for
the physical amplitude, some of them are fixed on a mass shell.  The
expansion on the light cone, so to say, ``separates the arguments", as a
result of which it becomes possible to study the physical UV asymptotic
behaviour by means of the expansion coefficients (when some momenta being
fixed on mass shell). As an important example, we can mention the evolution
equations for moments of QCD structure functions \cite{ap77}.  \vspace{2mm}

 The revealing of an {\it asymptotic freedom phenomenon} can be considered as
the most important result obtained in particle physics by the RG technique.
\medskip

  Historically, this discovery was made \cite{af} in the framework of the
SU(3) non--Abelian Yang-Mills model in the early 70s. Since that time this
model for the eight--component 4-vector field $B_{\mu}^{a b}(x)$ was adopted
as a basic ingredient for the QFT description of matter on the parton level.

  The key point is that self-interaction of this non-abelian gluonic
quantum field due to dominance of its unphysical components gives negative
contribution to the beta function perturbation expansion. For the two-loop
(scheme--independent) case
$$
\beta_{\rm QCD}(\alpha)= -\beta_1 \alpha^2- \beta_2 \alpha^3~ $$
with positive $\beta_{1,2}$ for a number $n_f$ of quark flavours small
enough.

   Correspondingly, the one-loop renorm-group expression
 $$
\bar\alpha_s^{(1)}(x;\alpha_s)=\frac{\alpha_s}{1+\alpha_s\beta_1\ln x}~,$$
for the QCD effective coupling exhibits a remarkable UV asymptotic behaviour
thanks to $\beta_1$ being positive. This expression implies, in contrast to
eq.(\ref{a1rg}), that the effective QCD coupling decreases as $x\sim Q^2$
increases and tends to zero in the UV limit. This feature discovered in the
early 70s, precisely corresponded to the parton physical picture of the
hadronic structure.   \vspace{2mm}

   One more interesting application of the RG method in the multicoupling
case, ascending to 50s \cite{ilya}, refers to special solutions, the
so-called separatrixes in a phase space of several invariant couplings. These
solutions relate effective couplings and represent scale-invariant
trajectories, like, e.g., $g_i=g_i(g_1)\,$ in the phase space which are
straight lines in the one-loop case.

 Some of them that are ``attractive" (or stable) in the UV limit, are related
to symmetries that reveal themselves in the high-energy domain. It was
conjectured that these trajectories may be related to {\it hidden symmetries
of a Lagrangian} and even could serve as a tool to find them. On this basis
the method was developed \cite{ks76} for finding out these symmetries. It was
shown that in the phase space of invariant couplings the internal symmetry
corresponds to a singular solution that remains a straight-line when higher
order corrections are taken into account. Such solutions corresponding to
supersymmetry were derived for some combinations of gauge, Yukawa and quartic
interactions.

  Generally, these singular solutions obey the relations
$$
\frac{dg_i}{dl}= \frac{dg_i}{dg_1}\frac{dg_1}{dl}\,,\;\,\, l=\ln x$$
which are known since Zimmermann's paper \cite{Z85} as {\it the reduction
equations}. In the 80s they were used \cite{OSZ} (see also review paper
\cite{z-rg} and references therein) in the UV analyzis of asymptotically free
models. Just for these cases the one-loop reduction relations are adequate to
physics.

  Quite recently some other application of this technique was obtained in
supersymmetric generalizations of Grand Unification scenario in the Standard
Model. It was shown \cite{EKT,K} that it is possible to achieve complete UV
finiteness of a theory if Yukawa couplings are related to the gauge ones in a
way corresponding to these special solutions, that is, to reduction
relations.  \medskip

   The mass-dependent technique described in Section \ref{ss3.4} was
successfully used for the development of the Dhar-Gupta approach
\cite{dhar83,dh-gu83} that led to finite perturbative predictions for a
physical quantity which is free of renormalization scheme ambiguities. In
paper \cite{gupta91}, this approach was reformulated for the mass-dependent
case with several coupling constants.  \medskip

  One more recent QFT development relevant to the renorm-group is the
``Analytic approach" to perturbative QCD (pQCD). It is based upon the
procedure of {\it Invariant Analyticization}\/ \cite{jinr96,np97} ascending
to the end of 50s. \par

 The approach consists in the combining of two ideas: the RG summation of UV
logs with analyticity in the $Q^2$ variable, imposed by spectral
representation of the K\"all\'en--Lehmann type which implements general
properties of the local QFT including the Bogoliubov condition of microscopic
causality. This combination was first devised \cite{bls59} to get rid of the
ghost pole in QED about forty years ago.  \par

 Here, the pQCD invariant coupling $\bar{\alpha}_s(Q^2)\,$ is transformed
into an ``analytic coupling" $\alpha_{\rm an}(Q^2/\Lambda^2)\equiv {\cal
A}(x)$ which, by construction, is free of ghost singularities due to
incorporating some nonperturbative structures.

 This analytic coupling ${\cal A}(x)$ has no unphysical singularities in the
complex $Q^2$-plane; its conventional perturbative expansion precisely
coincides with the usual perturbation one for $\bar{\alpha}_s(Q^2)\,$; it has
no extra parameters; it obeys a universal IR limiting value ${\cal A}(0)=
4\pi/\beta_0\,$ independent of the scale parameter $\Lambda$; it turns out to
be remarkably stable~\cite{np97} in the IR domain with respect to higher-loop
corrections and, in turn, to the scheme dependence. \par

  Meanwhile, the ``analyticized" perturbation expansion \cite{MSS97} for an
observable $F$, in contrast to the usual case, may contain specific functions
${\cal A}_n(x)$, instead of powers $\left({\cal A}(x)\right)^n\,$. In other
words, the pertubation series for $F(x)$, due to analyticity imperative, can
change its form in the IR region \cite{tmf99} turning into an asymptotic
expansion \`a la Erd\'elyi over a nonpower set $\{{\cal A}_n(x)\}\,$. \par

\section {RG expansion \label{s4}}                          

In 70s and 80s RGM was applied to (besides QFT) critical phenomena:
polymers, turbulence, non-coherent radiation transfer, dynamical chaos,
and so on.  Simpler and less sohpisticated motivation in critical
phenomena (than in QFT) makes this "explosion" of RG applications possible.

\subsection{Critical phenomena}

\subsubsection{\small\it Spin lattices}
 The so called renormalization group in critical phenomena is based on the
Kadanoff--Wilson procedure \cite{leo,ken} referred to as ``decimation" or
``blocking". Initially, it emerged from the problem of spin lattice. Imagine
a regular (two- or three- dimentional) lattice consisting of $N^d, ~d=2,3$
cites with an `elementary step' $a$ between them. Suppose, that at every site
a spin vector ${\bf\sigma}$ is sitting.  The Hamiltonian describing the spin
interaction of nearest neighbours
\beq
H = k \sum_i {\bf\sigma_i\cdot \sigma_{i \pm 1}} \eeq
contains $k$, the coupling constant. The statistical sum is obtained from
the partition function, $ S = < \exp (-H/\theta) >_{\rm aver.}.$

 To realize the blocking or decimation, one has to perform the ``spin
averaging" over block consisting of $n^d$ elementary sites. This is a very
essential step as far as it diminishes the degree of freedom number (from
$N^d$ to $(N/n)^d$). It destroys the small-range properties of the system
under consideration, in the averaging course some information being lost.
However, the long-range physics (like correlation length essential for phase
transition) is not affected by it, and we gain simplification of our problem.

 After this procedure, new effective spins ${\bf\Sigma}$ arise in sites of a
new effective lattice with a step $na$. We obtain also a new effective
Hamiltonian, with new effective coupling $K_n$ that has to be defined in the
averaging process as a function of $k$ and $n$
$$
H_{\rm eff}=K_n \sum_I {\bf\Sigma}_I\cdot {\bf \Sigma}_{I\pm 1}+\Delta H~,$$
where $\Delta H$ contains quartic and higher terms; $ \Delta H = \sum {\bf
\Sigma \cdot\Sigma ~~\Sigma \cdot\Sigma} + \cdots~.$

For the IR (long-distance) properties, $\Delta H$ is unessential. Hence, we
can conclude that the spin averaging leads to an approximate transformation,
\beq
 k \sum_i {\bf \sigma}\cdot {\bf\sigma} \to K_n {\bf\sum_I \Sigma\cdot
\Sigma} ~~,\eeq
 or, taking into account the ``elementary step" change, to
 $$
KW_n:~\left\{a \to n\,a, ~k \to K_n \right\}.  $$
 The latter is the  Kadanoff-Wilson transformation.

 In general, the ``new" coupling constant $K_n$ is a function of the ``old"
one and of the decimation index $n$. It is convenient to write it down in
the form $K_n= K(1/n,K)$. Then, the KW transformation can be formulated as
follows:
\beq\label{kw}
 KW(n): \left\{a \rightarrow n a, \quad k \to K_n =
K\left({1 \over n}, k\right)\right\} ~~. \eeq
These transformations obey the group composition law
$$  KW(n) \cdot KW(m) = KW(nm) $$ if
\beq
K(x, k) = K({x \over t}, K(t, k))\quad\left[x=
{1\over{nm}}, t= {1\over n}\right]~.\eeq       
This is just the RG symmetry.
\smallskip

 We observe the following points:
\begin{itemize}
\item The RG symmetry is approximate (due to neglecting by $\Delta H$).
\item The transformations $KW(n)\,$ are discrete.
\item There exist no reverse transformation to $KW(n)\,$.
\end{itemize}
 Hence, the `Kadanoff-Wilson renormalization group` is an {\it approximate
and discrete semi-group}. For a long distance (IR limit) physics, however,
$\Delta H$ is irrelevant, $\Delta(1/n)$ is close to continuum and it is
possible to use differential Lie equations.

 In application of these transformations to critical phenomena the notion of
a fixed point is important. As it was explained in Section \ref{ss3.1}, it is
usually associated with power-type asymptotic behavior. Note here that,
contrary to the QFT case considered in Section \ref{ss3.1}, in phase
transition physics we deal with  the IR stable point.

\subsubsection{\small\it Polymer theory}                            

 In the polymer physics one considers statistical properties of polymer
macromolecules which can be imagined as a very long chain of identical
elements. The number of elements $N$ could be as big as $10^5$, the
macromolecular size reaching several hundred Angstr\"oms.

 Such a big molecular chain forms a specific pattern resembling the pattern
of a random walk.  The central problem of the polymer theory is very close to
that of a random walk and can be formulated as follows.

 For a very long chain of $N$ ``steps" (the size of each step = $a$) one has
to find the ``chain size" $R_N$ as the distance between the ``start" and the
``finish" points, the distribution function of angles $\phi_i$ between
neighboring elements being given.

  The function $f(\phi)$ is defined by the forces between adjacent elements
depending on some external factors like temperature $T$.  The essential
feature of a polymer chain is the impossibility of a self-intersection. This
is known as an {\it excluded volume} effect in the random walk problem. In
reality, polymer molecules are swimming in a solvent and form {\it globulars}.

  For large $N$ values the molecular size $R_N$ follows the power Fleury law
$R_N \sim N^\nu $ with $\nu$, the Fleury index.  When $N$ is given, $R_N$ is
 a functional of $f(\phi)$ which depends on external conditions (e.g.,
temperature $T$, properties of solvent, \etc).  If $T\/$ increases, $R_N\/$
increases and at some moment globulars touch one another. This is the
polymerization process very similar to a phase transition phenomenon.

 The Kadanov--Wilson RG (KWRG) blocking ideology has been used in polymer
physics by De Gennes \cite{degen}. The key idea is a grouping of $n$ chain
 subsequent elements into a new ``elementary block". This grouping operation
is very close to Kadanoff's blocking. It leads to the transformation
$$
\left\{ 1 \to n ~; ~~ a \to A_n \right\}~ $$
which is analogous to one for spin lattice decimation. This
transformation must be specified by a direct calculation which gives the
explicit form of $A_n=\bar{a}(n,a)$. Here we have a discrete semi-group.
Then, by using the KWRG technique, one finds the fixed point, obtains the
Fleury power law and can calculate its index $\nu$.

  Generally, the excluded volume effect yields some complications.  However,
inside the QFT RG framework it can be treated rather simply \cite{alhim} by
introducing an additional argument similar to finite length $L$ in transfer
problem and particle mass $m$ in QTF.  \smallskip

Besides polymers, the  KWRG approach has been used in some fields of physics,
like percolation, noncoherent radiation transfer \cite{bell}, dynamical
chaos \cite{chaos} and some others.
\smallskip

Meanwhile, the original QFT--RG approach proliferated into the theory of
turbulence.

\subsubsection{\small\it Turbulence \label{sss4.1.3}}

   To formulate the turbulence problem on the ``RG language" one has to
perform the following steps ~\cite{dom,adj,vas}:
\begin{enumerate}
\item Introduce the generating functional for correlation functions.
\item Write the path integral representation for this functional.
\item By changing the  function integration variable find the equivalence
     of the classical statistical system to some quantum field theory model.
\item Construct the system of Schwinger--Dyson equations for this equivalent
     QFT.
\item Perform the finite renormalization procedure.
\item Derive the RG equations.
\end{enumerate}

\subsection{Paths of RG expansion}

 RG is expanded in diverse fields of physics in two different ways:
\begin{itemize}
\item by  direct analogy with the Kadanov-Wilson construction (averaging
over some set of degrees of freedom) in polymers, non-coherent transfer and
percolation, i.e., constructing a set of models for a given physical problem.
\item search for an exact RG symmetry by proof of the equivalence with a QFT
model: e.g., in turbulence (Refs.~\cite{dom,vas}), plasma turbulence
\cite{pell} and some others.
\end{itemize}
\bigskip

 To the question {\sf Are there different renormalization groups?}  the
 answer is positive:
\begin{enumerate}
\item In QFT and some simple macroscopic examples (like, one--dimentional
transfer problem) , ~RG~ symmetry is an exact symmetry of the solution
formulated in its natural variables.
\item In turbulence, continuous spin-field models and some others, it
is a symmetry of an equivalent QFT model.
\item In polymers, percolation, \etc, (with KW blocking), the RG
transformation is a {\it transformation between different auxiliary models}
(specially constructed for this purpose) of a given system.
\end{enumerate}
\medskip

 As we have shown, there is no essential difference in the mathematical
formalism. There exists, however, a profound difference in physics:
\par --- In  cases 1 and 2 (as well as in some macroscopic examples), the RG
is an exact symmetry of a solution. \par
 --- In the Kadanov--Wilson--type problem (spin lattice, polymers,
\etc), one has to construct a set ${\cal M}$ of models $M_i$. The KWRG
transformation
\beq
R(n) M_i = M_{n i} ~,~~\mbox{with integer} ~n \eeq
{\it is acting inside a set of models}.

\subsection{Two faces of RG in QFT}                                 

 As it was explained in Section 1.4, the vacuum, i.e., the interparticle
space, contain vacuum fluctuations. Due to them, the charge of a particle
is screened. In accordance with Dirac eq.(\ref{dir}), in momentum space the
$Q^2$ dependence of an electron charge can be presented
\beq\label{e-dir}
e(Q^2) =e \left\lbrace 1+{\alpha \over 6\pi}\ln (Q^2 r_e^2)
+ \dots \right\rbrace ~ ; ~e^2= e^2(1/r_e^2)= 1/137. \eeq    
in terms of the classical electron charge and of electron Compton
lenght. \par

  The first idea of an additional symmetry in this problem was born by
St{\"u}eckelberg and Peterman \cite{stp}. In their pioneering investigation
the very existence of group transformation was discovered within the
renormalization procedure the result of which contains finite arbitrariness.
Just this degree of freedom in finite renormalized expressions was used by
Bogoliubov and Shirkov in Refs.\cite{bs-55a}---\cite{nc-56}. Roughly
speaking, this corresponds to the change $ r_e \to 1/\mu$.

 The basic idea was that, instead of $1/r_e$, one can use some other
reference point $\mu$. This is equivalent to introducing of {\it a new degree
of freedom} associated with the reference point scale. Instead of
(\ref{e-dir}) we have
\beq\label{bsh}
e(Q^2/\mu^2)=e_{\mu}\left\lbrace 1+{\alpha_{\mu} \over 6\pi}
\ln{Q^2\over\mu^2} + \dots  \right\rbrace ~~. \eeq                           

 Here, the effective charge is considered {\it after} the subtracting of
infinities and is given by a ``finite representation" (\ref{bsh}). The
RG symmetry is formulated in terms of the $Q^2$ scale and $\mu$ represents
the reference point.

 Another approach was used by Gell-Mann and Low \cite{gml}. Their paper was
devoted to the short distance behaviour in a nonlocal QED with a cutoff
$\Lambda$, and the ``$\Lambda$ degree of freedom" was used to analyze the
UV behaviour.  Instead of renormalization, there is a regularization and the
charge is given by the ``singular representation"
\beq
 e(\Lambda) =e\left(1+{\alpha\over 6\pi}\ln \Lambda^2 r_e^2 +
\cdots~\right)  \eeq                  
which is singular in the limit $\Lambda \to \infty$.

 We can draw a transparent picture (as was commented later by Wilson in
his Nobel lecture) of the last approach. Imagine an electron
of a finite size, smeared over a small volume with the radius $R_i=
\hbar/c\Lambda_i~$, $\ln(\Lambda^2/m^2_e)\gg 1 $. The electric charge $e_i$
of such a non-local electron is considered as depending on the cut-off
momentum $\Lambda_i$ so that this dependence accumulates the vacuum
polarization effects which, in reality, take place at distances from the
point electron smaller than $R_i$. We deal with a set of models of the
non-local electron correponding to different values of the cut-off
$\Lambda_i$.
 Here, $e_i$ depends on $R_i$ and the vacuum polarization effects in the
excluded volume $R_i^3$ should be subtracted. In this language, the RG
transformation is the transition from one value of the smearing radius to
another $R_i \to R_j$, simultaneously with a corresponding change of the
effective electron charge $e_i \to  e_j$. In other words, the RG symmetry
here is that related to operations in the space of models of non-local QED
constructed in such a way  that at large distances every model is equivalent
to the real local one.

\section{RG symmetry in mathematical physics \label{s5}}         
 \subsection{Functional self-similarity \label{ss5.1}}

The RG transformations discussed above have  close connection with
the concept of a self-similarity(SS).
The SS transformations for problems formulated by nonlinear partial DEs
are well known, since the last century, mainly in dynamics of liquids
and gases. They are one parameter $\lambda$ transformations
defined as simultaneous power scaling of independent variables
$~z=\{x,t,\ldots\}~$, solutions $~f_k(z)~$ and other functions $~V_i(z)~$
$$
S_\lambda : ~~\left\{ x\to x\lambda~, t\to t\lambda^a~,~f_k(z)\to
f_k^\prime(z^{\prime}) =\lambda^{\varphi_k} f_k(z^{\prime})~,~
~V_i(z) \to =\lambda^{\nu_i} V_i(z^{\prime})~\right\} ~~ $$
entering into the equations.
\par To emphasize their power structure, we use a term {\it power}
self-similarity = PSS. According to Zel'dovich and Barenblatt,
\cite{zeld,baren} the PSS can be classified as: \par

a/ {\it PSS of the 1st kind} \\
with all indices $a, ... \varphi, \nu , ...$
being (half)integers (Integer PSS) that are usually found from
the theory of dimensions; \par
b/ {\it PSS of the 2nd kind} \\
with irrational indices (Fractal PSS) which should be defined from dynamics.

To relate RG with PSS, let us turn to the solution of the renorm-group FE
$$
\gbar (xt, g) = \gbar (x, \gbar (t, g))~. $$
Its general solution is known; it depends on an arbitrary function of
one argument -- see eq.(\ref{2-13}). However, at the moment we are
interested in a special solution linear in the second argument:
$\gbar (x, g)=gf(x).$ The function $f(x)$ should satisfy the equation
$~f(xt)=f(x)f(t)~$ with the solution $~f(x)=x^{\nu}~$. Hence,
$\gbar (x, t) = g x^\nu $.
This means that in our special case, linear in $g$, the RG transformation
(\ref{1-10}) is reduced to PSS transformation,
\beq \label{6-1}
R_t \Rightarrow  \{ x \to xt^{-1}, ~g \to g t^\nu \} = S_t ~~.\eeq    %

 Generally, in RG, instead of a power law, we have arbitrary functional
dependence. Thus, one can consider transformations (\ref{1-10}), (\ref{1-13})
and (\ref{1-17}) as functional generalizations of usual (i.e., power)
self-similarity transformations. Hence, it is natural to refer to them as to
the transformations of {\it functional scaling} or functional
(self)similarity (FS) rather than to RG-transformations. In short,
$$
{\rm RG} \equiv {\rm FS} ~~, $$
with FS standing for {\sf Functional Similarity}.

 We can now answer the question concerning the physical meaning of the
symmetry underlying FS and the Bogoliubov's renorm--group. As we have
mentioned, it is not a symmetry of the physical system or the equations of
the problem at hand, but a {\it symmetry of a solution} considered as a
function of the relevant physical variables and suitable boundary conditions.
A symmetry like that can be related, in particular, to the invariance of a
physical quantity described by this solution with respect to the way in which
the boundary conditions are imposed. The changing of this way constitutes a
group operation in the sense that the group composition law is related to the
transitivity property of such changes.

 Homogeneity is an important feature of the physical systems under
consideration. However, homogeneity can be violated in a discrete manner.
Imagine that such a discrete inhomogeneity is connected with a certain value
of $\,x\,$, say, $\,x=y$. In this case the RG transformation with the
canonical parameter $\,t\,$ will have the form (\ref{1-12}) with the group
composition law (\ref{1-13}).

 The symmetry connected with FS is a very simple and frequently encountered
property of physical phenomena.  It can easily be ``discovered" in numerous
problems of theoretical physics like classical mechanics, transfer theory,
classical hydrodynamics, and so on \cite{O-7,mamik,kiev84}.

\subsection{Recent application to boundary value problem\label{ss-bvp}}

Recently, some interesting attempts have been made to {\it use the RG concept
in classical mathematical physics}, in particular, to study strong nonlinear
regimes and to investigate asymptotic behavior of physical systems described
by nonlinear PDEs.

 About a decade ago, the RG ideas were applied by late Veniamin Pustovalov
with co-authors \cite{KP-87-90} to analyze a problem of generating higher
harmonics in plasma. This problem, after some simplification, was reduced to
a couple of partial DEs with the boundary parameter -- ``solution
characteristic" -- explicitly included. It was proved that corresponding
solutions admitted an exact symmetry group that takes into account
transformations of this boundary parameter, which is related to the amplitude
of the magnetic field at a critical density point. The solution symmetry
obtained was then used to evaluate the efficiency of harmonics generation in
cold and hot plasma. The advantageous use of the RG-approach in solving the
above particular problem gave promise that it may work in other cases and
this was illustrated in \cite{KKP_RG_91} by a series of examples for various
boundary value problems.

 Moreover, in Refs. \cite{O-7,KKP_RG_91} the possibility of devising a
regular method for finding a special class of symmetries of solution to the
boundary value problem (BVP) in mathematical physics, namely, RG-type
symmetries, was discussed. The latter are defined as solution symmetries with
respect to transformations involving parameters that enter through the
equations as well as through the boundary conditions in addition to (or even
rather than) the natural variables of the equations.

 As it is well known, the aim of the modern group analysis \cite{Oves,Ibr},
which goes back to works by S. Lie \cite{Lie}, is to find symmetries of DEs.
This approach does not include a similar problem of studying the symmetries
of solutions of these equations. Outside the main direction of both the
classical and modern analysis, there remains as well a study of solution
symmetries with respect to transformations involving not only the variables
present in the equations, but also parameters entering into the solutions
from boundary conditions.

 From the afore-said it is clear that the symmetries which attracted
attention in the 50s in connection with the discovery of the RG in QFT were
those involving the parameters of the system in the group transformations. It
is natural to refer to these symmetries related to {\it FS (or RG-type)
symmetries}.  \smallskip

 It should be noted that the procedure of revealing the FS symmetry (FSS),
or some group feature, similar to the FS regularity, in any partial case
(QFT, spin lattice, polymers, turbulence and so on) up to now is not a
regular one. In practice, it needs some imagination and atypical manipulation
``invented" for every particular case --- see the discussion in \cite{umn94}.
By this reason, the possibility to find a regular approach to constructing
FSS is of principal interest.
\smallskip

  Recently, a possible scheme of this kind was presented as applied to a
mathematical model that is described by a BVP. The leading idea
\cite{O-7,KKP_RG_91,Shr_Pisa-95} in this case is based on the fact that
solution symmetry for this system can be found in a regular manner by using
the well-developed methods of modern group analysis.

 The scheme that describes devising of FSS and its application is then
formulated \cite{kov-RG96,KPSh_JMP_98} as follows.  Firstly, a specific
RG-manifold should be constructed. Secondly, some auxiliary symmetry, i.e.,
the most general symmetry group admitted by this manifold is to be found.
Thirdly, this symmetry should be restricted on a particular solution to get
the FSS. Fourthly, the FSS allows one to improve an approximate solution or,
in some cases, to get an exact solution.  \smallskip

  Depending on both a mathematical model and boundary conditions, the {\it
first step} of this procedure can be realized in different ways. In some
cases, the desired FS-manifold is obtained by including parameters, entering
into a solution via an equation(s) and a boundary condition, in the list of
independent variables. The extension of the space of variables involved in
group transformations, e.g., by taking into account the dependence of
coordinates of the renorm--group operator upon differential and/or non-local
variables (which leads to the Lie---B\"acklund and non-local transformation
groups \cite{Ibr}) can also be used for constructing the FS--manifold. The
use of the Ambartsumian invariant embedding method \cite{Ambar} and of
differential constraints sometimes allows reformulations of a boundary
condition in a form of additional DE(s) and enables one to construct the
FS-manifold as a combination of original and embedding equations
(or differential constraints) which are compatible with these equations. At
last, of particular interest is the perturbation method of constructing the
FS--manifold which is based on the presence of a small parameter. \par
\smallskip

 The {\it second step}, the calculating of a most general group $\cal{G}$
admitted by the FS--manifold, is a standard procedure in the group analysis
and has been described in detail in many texts and monographs -- see, for
example, \cite{Oves,Olver}.
\smallskip

 The symmetry group $\cal{G}$ thus constructed cannot as yet be referred to
as a renorm--group. In order to obtain this, the next, {\it third step}
should be done which consists in restricting $\cal{G}$ on a solution of a
boundary value problem.  This procedure utilizes the invariance conditionš
and mathematically appears as a ``combining" of different coordinates of
group generators admitted by the FS--manifold.
\smallskip

 The {\it final step}, i.e., constructing analytic expression for the
solution of the boundary value problem on the basis of the FS, usually
presents no specific problems.

A review of the results, which were obtained on the basis of the formulated
scheme, can be found, for example, in \cite{KPSh_JMP_98,K_MGA_96,ks-tmp99}.

  We mention briefly, the FS analysis result for a particular problem of
nonlinear optics, the problem of the laser beam self-focusing in a nonlinear
medium. Here, one have a BVP for a coupled system of two nonlinear
PDEs with the boundary condition given in a form of two one-argument
functions. With help of RG=FS approach one new exact analytic and one new
approximate analytic solution (for the practically important Gaussian initial
transverse profile) has been found \cite{optics}.     \par
  The important qualitative features of this example are: \par
\begin{itemize}
\item[--] the {\it two-dimension structure} has been analysed, that
has a singularity in the behavior of derivatives with respect to
transverse coordinate, 
\item[--] the algebraic structure of the
FSS operators is different from that of ``usual RG of the QFT
type". Here, we meet with a
set of {\it several infinitesimal renormgroup operators}, each of
those can be used to reconstruct the analytic (exact or
approximate) solution of BVP starting from the perturbative theory
result. Moreover, renormgroup operators for exact solutions of BVP
obtained appear as Lie-B\"acklund ({\it not} Lie) infinitesimal operators.
\end{itemize}
 \medskip

  Up to now the outlined regular method is feasible for systems that can be
described by DEs and is based on the formalism of modern group analysis.
However, it seems also possible to extend this approach to boundary value
problems that are not described just by differential equations. A chance of
such an extension is based on recent advances in group analysis of systems of
integro-differential equations~\cite{KKP_DE_93} which allow transformations
of both dynamical variables and functionals of a solution to be formulated
\cite{KKP_JNMP_96}. More intriguing is the issue of a possibility of
constructing a regular approach for more complicated systems, in particular
to those having an infinite number of degrees of freedom.  The formers can be
represented in a compact form by functional (or path) integrals. %
 \medskip

{\large\bf Acknowledgments }
\smallskip

 The author is grateful to Professor Ashoke Mitra for invitation to
participate in this book. He is indebted to D.V. Kazakov, V.F. Kovalev, and
I.L. Solovtsov for useful discussion and comments.  This work was partially
supported by grants of Russian Foundation for Fundamental Research (RFFR
projects Nos 96-15-96030 and 99-01-00091) and by INTAS grant No 96-0842.

\end{document}